\documentclass{jfm}
\usepackage{graphicx}
\usepackage{epstopdf, epsfig}
\usepackage{amsmath,amsfonts}

\shorttitle{Resonant triads in two layer shallow water}
\shortauthor{A.Owen, R. Grimshaw, B. Wingate}

\title{Fast and slow resonant triads in the two layer rotating shallow water equations}

\author{Alex Owen\aff{1}
  \corresp{\email{ao306@exeter.ac.uk}},
  Roger Grimshaw\aff{1}
 \and Beth Wingate \aff{1}}

\affiliation{\aff{1}College of Engineering, Mathematics and Physical Sciences, University of Exeter, North Park Road, 
Exeter EX4 4QF, UK}

\begin{document}

\maketitle

\begin{abstract}
In this paper we examine triad resonances in a rotating shallow water system when there are two free interfaces. This allows for an examination in a relatively simple model of the interplay between baroclinic and barotropic dynamics in a context where there is also a geostrophic mode.  In contrast to the much-studied one-layer rotating shallow water system, we find that as well as the usual slow geostrophic mode, there are now two fast waves, a barotropic mode and a baroclinic mode. This feature permits triad resonances to occur between three fast waves, with a mixture of barotropic and baroclinic modes, an aspect which cannot occur in the one-layer system. There are now also two branches of the slow geostrophic mode with a repeated branch of the dispersion relation. The consequences are explored in a derivation of the full set of triad interaction equations,  using a multi-scale asymptotic expansion based on a small amplitude parameter. The derived nonlinear interaction coefficients are confirmed using energy and enstrophy conservation. These triad interaction equations are explored with an emphasis on the parameter regime with small Rossby and Froude numbers. 
\end{abstract}

\begin{keywords}
Authors should not enter keywords on the manuscript, as these must be chosen by the author during the online submission process and will then be added during the typesetting process (see http://journals.cambridge.org/data/\linebreak[3]relatedlink/jfm-\linebreak[3]keywords.pdf for the full list)
\end{keywords}
\section{Introduction}
The one layer rotating shallow water equations are well studied equations in the context of the interaction of fast gravity and slow quasigeostrophic components of the flow. In this paper we extend this work to the case with two free interfaces. This is then a simplest model of interaction between barotropic and baroclinic modes. We show that there are significant differences when a second free layer is introduced: new triad resonances exist between the two vertical modes that are not present in the one layer equations. This is also in contrast to the equations with the top layer held rigid. Unlike previous work, the resonances exhibit qualitative changes in behaviour dependent on the strength of rotation effect. The focus is on the behaviour in the weakly nonlinear limit through the multiple scales method.

Although triad resonances had been explored in other areas such as solid state physics (see \cite{ziman1960electrons} for example), in geophysical fluid dynamics interest in resonant wave interactions began with \cite{phillips1960dynamics} in a study of water waves, where in fact triad interactions are not allowed and instead quartet interactions dominate. Over the next decade several papers followed applying the method to different situations where triad resonances are supported:  \cite{ball1964energy} applied the ideas to two layer non-rotating systems, and \cite{mcgoldrick1965resonant} for capillary waves, both of which have a suitable dispersion relation for the interaction of three waves. Other related work that expanded knowledge of different time scales included: \cite{hasselmann1962non} who furthered the surface wave work,  \cite{benney1967propagation} who summarised the method and \cite{newell1969rossby} who applied it to Rossby wave packets. A review paper by \cite{phillips1981wave} covers the historical development of the area, and general details can also be found in \cite{craik1988wave}. Of these papers the closest to the current work is that of Ball, who analysed the triads in a two layer scenario for the case of no rotation. However rotation is not considered negligible in most geophysical applications: in this paper we include it, introducing the geostrophic modes.

More recently, work on multi-scale shallow water theory has been done by \cite{warn1986statistical}, \cite{babin1997regularity}, and by \cite{embid1996averaging} whose approach is used in parts of this paper. In \cite{embid1996averaging} the one layer shallow water equations were approached using the parameter limit from quasigeostrophy but retaining a fast time scale. It was found that the dynamics split into an equation of motion describing not just the well-known quasigeostrophic approximation for the inertial part, but also a second equation, coupled to the first, describing the gravity waves (on the fast time scale) interacting with the geostrophic part. This interaction was shown to be one-way: the inertial part is unaffected by the gravity waves and acts as a catalyst to the gravity wave interactions (for detail see \cite{ward2010scattering}).

A series of papers by Reznik, Zeitlin and collaborators have examined the wave interactions of layered fluid models. We note especially \cite{reznik2001nonlinear} which explored different geostrophic limits in the one layer shallow water equations with compact support assumed for the initial conditions, and \cite{zeitlin2003nonlinear} who considered a two layer shallow water model in the rigid lid limit. They found that the long time evolution of the slow part of the flow is unaffected by the fast part as the infinite domain allowed the fast modes to disperse. These boundary conditions give a very different scenario to this present work in a variety of ways: the approach did not rely on resonant conditions, had a non-periodic domain and the rigid lid condition removes the second free surface. Recently  \cite{thomas2016resonant} re-examined the one layer model, and found that a restriction to a periodic domain might allow a continuing interaction between the fast and slow modes. We also note that \cite{zeitlin2013resonant} investigated a similar two layer model but in the half-plane case where there is a boundary along which Kelvin waves can propagate.

Interest in the interaction between fast and slow time scales in the context of climate dynamics began with two key papers by \cite{hasselmann1976stochastic} and \cite{frankignoul1977stochastic} suggested that observed variation in the climate could be explained by modelling of the climate with a small scale stochastic forcing behaving as \lq{}weather\rq{}. Follow up work in \cite{frankignoul1985sea} and \cite{frankignoul1998air} extended this model to consider the effect of the sea-surface. Soon after this last investigation \cite{goodman1999model} developed an `active coupling model' to investigate the mechanism of growth on slow decadal time scales. This involves modelling the interaction between two active layers with simulated dynamics, as opposed to having a passive atmosphere and dynamic ocean. Later work by \cite{farneti2007coupled} showed results in a numerical climate model close to those predicted by the coupling model, using an active upper layer. These papers build up a picture of climate modelling requiring contributions from many scales and particularly highlight the importance of interactions between layers: there is a need for work that investigates these effects. In this paper we are motivated by these concepts to consider the simplest possible model of interaction between separate free layers over many scales and time frames. Particularly we contrast this to the single layer and rigid lid  systems, where either the baroclinic or barotropic mode is absent.

In addition, this paper is intended to both extend the two layer work of \cite{ball1964energy} to include rotation and to extend the rotating one layer work such as that done by \cite{embid1996averaging} to include a second layer. These two properties have not been explored together, as the rigid lid case is often adopted. It is important to understand the implications on the behaviour of the system when the rigid lid assumption is dropped to be sure that relevant effects are not being excluded. In this paper it is shown that there is new behaviour in the interactions of the gravity waves, behaviour that only exists where there are two free layers. In addition we show that the rotation of the system has an important effect on these interactions, altering the range of wavenumbers they affect and even their existence. In certain regimes these interactions are particularly notable for being restricted to a small range of angles of incidence between waves as well as being restricted to interactions between waves of large spatial scale differences.

In section \ref{bas_eq} the basic equations are introduced and properties of the system are expanded-upon. In sections \ref{qg_lim}, and \ref{non-lin_coeff} we conduct a multiple scales analysis for the two layer equations up to the first closure and we find the nonlinear interaction coefficient explicitly for the different wave modes in the system. We then show that the geostrophic part of the flow is in fact equivalent to the standard two layer quasigeostrophic equations.

A second method of analysis that places more emphasis on the conservation laws is conducted (this can be found in \cite{grimshaw2007nonlinear} for example). This brings additional insight to the problem in section \ref{vanmeth} of this paper. 

We make links between the two layer system and its common simplification to the rigid lid system (see \cite{zeitlin2003nonlinear} for example). We see a difference in behaviour between the two systems: when no rigid lid assumption is made, there is coupling of internal and external wavemodes (found in the non-rotating case in \cite{ball1964energy}). Section \ref{fff_sec} explicitly explores this link by consideration of the possible resonant triads.

\section{Formulation of the basic equations} 
\label{bas_eq}
The two layer rotating shallow water equations in standard form and notation, see \cite{salmon1998lectures} for instance, are:

In the top layer:
\begin{align} 
\frac{D\boldsymbol{u_1}}{Dt}&+f\boldsymbol{\hat{z}}\times\boldsymbol{u_1}=-\nabla p_1 ,\label{mom1}\\
\frac{\partial h_1}{\partial t}&+\nabla\cdot(h_1 \boldsymbol{u_1})=0,\\
p_1&=g(\eta_1+\eta_2). \label{eta1}
\end{align}

\begin{figure}
\centerline{\includegraphics[width=10cm]{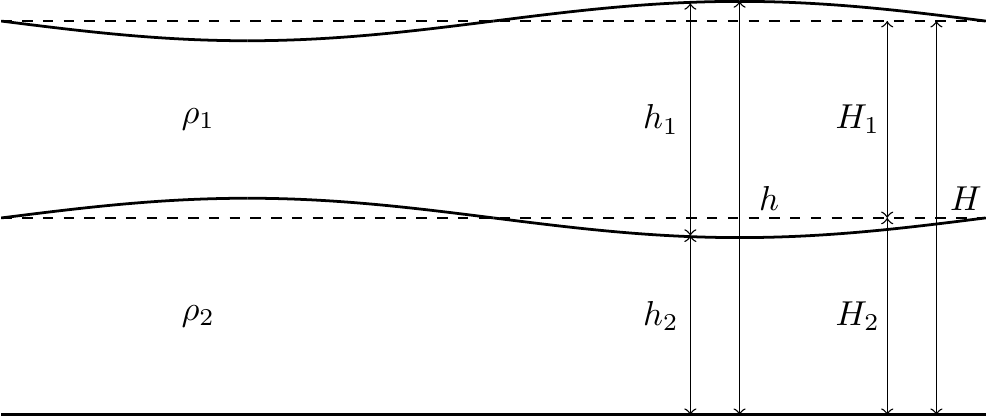}}
\caption{Diagram to show the definitions of the two layers.}
\label{two_layer_diagram}
\end{figure}

In the bottom layer:
\begin{align}
\frac{D\boldsymbol{u_2}}{Dt}&+f\boldsymbol{\hat{z}}\times\boldsymbol{u_2}=-\nabla p_2 , \label{mom2}\\ 
\frac{Dh_2}{Dt}&+\nabla\cdot(h_2 \boldsymbol{u_2})=0,\\
p_2=g(\eta_1+\eta_2)&-\left(\frac{\rho_2-\rho_1}{\rho_2}\right)g \eta_1=g\left(\frac{\rho_1}{\rho_2}\eta_1+\eta_2\right). \label{eta2}
\end{align}
The subscripts 1, 2 refer to the top and bottom layers respectively for $\boldsymbol{u}_i$ the velocity and $h_i$ the layer height. Figure \ref{two_layer_diagram} shows the set up of the system. Note that the displacements $\eta_i=h_i-H_i$ are not marked. The equations are only coupled through the pressure terms $p$.

We non-dimensionalise by assuming that the variables in each layer are the same order of magnitude as follows:
\begin{align} \nonumber
&x,y\sim L, \qquad u_i,v_i\sim U, \qquad t\sim L/U,\\
&H_i\sim H, \qquad \ \eta_i\sim D, \qquad \ \ \ p_i\sim gD. 
\end{align}
The length scale $L$ can be chosen to be one of the two deformation scales $f/c_m$ where $c_m$ is the linear long wave phase speed defined below in (\ref{c_calc}).
We reduce the parameters to the standard non-dimensional set of Rossby, Froude, and amplitude ratio respectively:
\begin{align}
Ro=\frac{U}{fL}, \qquad Fr=\frac{U}{\sqrt{gH}}, \qquad \theta=\frac{D}{H}.
\end{align}
And the resulting non-dimensional equations are:
\begin{eqnarray}
\frac{D\boldsymbol{u}'_i}{Dt'}+Ro^{-1}\boldsymbol{\hat{z}}\times\boldsymbol{u}_i\rq{}=-Fr^{-2}\theta\nabla p'_i,\\
\frac{\partial \eta'_i}{\partial t'}+\theta^{-1}\nabla\cdot\boldsymbol{u}'_i+\nabla\cdot(\eta'_i \boldsymbol{u}'_i)=0,
\end{eqnarray}
where $'$ indicates the non-dimensionalised variables. In the asymptotic analysis we will take the same small parameter assumption that forms part of the quasigeostrophic limit: this is the assumption $Fr,Ro,\theta \sim\epsilon$ with $0<\epsilon\ll1$. This leads to the non-local form in (\ref{non_loc1}). In order to consider different physical scenarios the analysis will continue with the original dimensional variables. However there is an underlying assumption that implicitly this is an asymptotic limit as $\epsilon\rightarrow 0$.

If we write the variables in the vector form:
\begin{eqnarray}
\boldsymbol{u}= 
\left[ \begin {array}{c} \boldsymbol{u}_1\\ \noalign{\medskip}\eta_1\\ \noalign{\medskip}\boldsymbol{u}_2\\ \noalign{\medskip}\eta_2 \end {array} \right]=
 \left[ \begin {array}{c} u_1\\  \noalign{\medskip}v_1\\ \noalign{\medskip}\eta_1\\ \noalign{\medskip}u_2\\ \noalign{\medskip} v_2\\ \noalign{\medskip}  \eta_2 \end {array} \right], 
\end{eqnarray}
then we can write the system of equations linearised around layers of heights $H_1$, $H_2$ as:
\begin{eqnarray}
\frac{\partial \boldsymbol{u}}{\partial t}+\frac{1}{\epsilon}\mathcal{L}(\boldsymbol{u})+\mathcal{N}(\boldsymbol{u},\boldsymbol{u})=0, \label{non_loc1}
\end{eqnarray}
where $\mathcal{L}$ is the linear operator and  $\mathcal{N}$ is the nonlinear part. We define the two-dimensional spatial Fourier transform with wavenumbers $\boldsymbol{k}=(k,l)$ as follows:
\begin{align}
\mathcal{F}[a(\boldsymbol{x})]=\hat{a}(\boldsymbol{k})=\int_{-\infty}^{\infty}a(\boldsymbol{x})e^{i\boldsymbol{k}\cdot\boldsymbol{x}}d\boldsymbol{x},
\end{align}
which gives the following property of differentiation:
\begin{align}
\mathcal{F}\left[\frac{\partial a(\boldsymbol{x})}{\partial x_j}\right]=ik_j\hat{a}(\boldsymbol{k}) .
\end{align}
It should be noted that we use the full Fourier transform, as through the paper we will consider both infinite domains and periodic domains: the periodic domain simply being the restriction of the Fourier transform to a discrete subset of the wavenumbers.

Taking the Fourier transform of (\ref{non_loc1}) we get the following matrix operator:
\begin{eqnarray}
\mathcal{L}=
\left[ \begin {array}{cccccc} 0&-f&igk&0&0&igk
\\ \noalign{\medskip}f&0&igl&0&0&igl\\ \noalign{\medskip}iH_1k&iH_1l&0&0&0&0\\ \noalign{\medskip}0&0&ir_\rho gk&0&-f&igk\\ \noalign{\medskip}0&0&ir_\rho gl&f&0&igl
\\ \noalign{\medskip}0&0&0&i H_2k&iH_2l&0
\end {array} \right].
\end{eqnarray}
Here $r_\rho={\rho_1}/{\rho_2}$.

In the next section (\ref{vert_bas}) we transform to skew-Hermitian form, to directly find an orthogonal basis of eigenvectors.
\subsection{Transformation of the linear operator into the vertical mode basis}
\label{vert_bas}
As is commonly done in geophysical fluid dynamics (see \cite{gill1982atmosphere} or \cite{vallis2006atmospheric} for example) we decompose into a vertical mode basis which we use in much of the rest of the paper.

The appropriate transformation (for example see \cite{ball1964energy}) can be found to be:
\begin{subequations}
\label{both_trans}
\begin{align} \label{transf}
\boldsymbol{u}_ m&=L_{ m}H_1\boldsymbol{u}_1+H_2\boldsymbol{u}_2,\\
p_ m&=\frac{1}{c_ m}(L_{ m}H_1p_1+H_2p_2),
\end{align}
\end{subequations}
where:
\begin{align}
c_ m^2=g\frac{H_1+H_2}{2}&+ m g\sqrt{\left(\frac{H_1-H_2}{2}\right)^2+H_1H_2r_\rho}, \label{c_calc}\\
L_ m H_1=\frac{H_1-H_2}{2}&+ m\sqrt{\left(\frac{H_1-H_2}{2}\right)^2+H_1H_2r_\rho}.\\
 m&=\pm \nonumber
\end{align}
$ m$ defines the vertical mode; in this form the system is split into modes $ m=+,-$ instead of the two layers $i=1,2$ as previously. 

Notice that the transformation operates on the pressure in each layer, not the perturbation heights $\eta_i$. More simply using (\ref{eta1}) and (\ref{eta2}) we can write the direct transform from $\eta_i$:
\begin{align}
p_ m=gL_ m c_ m\eta_1+g\eta_2
\end{align}
This transforms the linear operator into the following in Fourier space:
\begin{eqnarray}
\mathcal{L}=
 \left[ \begin {array}{cccccc} 0&-f&c_+ki&0&0&0
\\ \noalign{\medskip}f&0&c_+li&0&0&0
\\ \noalign{\medskip}c_+ki&c_+li&0&0&0&0
\\ \noalign{\medskip}0&0&0&0&-f&c_-ki
\\ \noalign{\medskip}0&0&0&f&0&c_-li 
\\ \noalign{\medskip}0&0&0&c_-ki&c_-li&0
\\ \end {array} \right] ,
\label{mode_lin}
\end{eqnarray}
It can be seen from (\ref{mode_lin}) that the two modes are entirely decoupled in the linear part: this means all coupling will now appear in the nonlinear interactions. Throughout the rest of this paper the modes are referred to as internal ($ m=-$) and external ($ m=+$). We adopted this convention in reference to the rigid lid limit (discussed in section \ref{rig_lid}, see also \cite{salmon1998lectures} for example). 

The frequencies and modal functions correspond to the eigenvalues and eigenvectors of the matrix. The frequencies are 
\begin{align} \nonumber
\omega=
&-\sqrt{c_{+}^2|\boldsymbol{k}|^2+f^2},\ 0,\ \sqrt{c_{+}^2|\boldsymbol{k}|^2+f^2},\\
&-\sqrt{c_{-}^2|\boldsymbol{k}|^2+f^2},\ 0,\ \sqrt{c_{-}^2|\boldsymbol{k}|^2+f^2} \addtocounter{equation}{1}\tag{\theequation \textit{a-f}}
\label{disprel_solns}
\end{align}
and in this basis the corresponding (orthonormal) modal functions are respectively:
\begin{align} \nonumber
 &\left[ \begin {array}{c} {\frac { \left( -k\omega+ ifl \right) }{\sqrt {2} \left| k \right| \omega}}\\ 
 \noalign{\medskip}{\frac { \left( - ifk- l\omega \right) }{\sqrt {2} \left|k \right| \omega}}\\ 
 \noalign{\medskip}{\frac { \left| k \right|  c_+}{\sqrt {2}\omega}}\\ 
 \noalign{\medskip}0\\ \noalign{\medskip}0\\ \noalign{\medskip}0\end {array} \right] 
,
 \left[ \begin {array}{c} -{\frac { il c_+}{\omega}}\\ 
\noalign{\medskip}{\frac { ik c_+}{\omega}}\\ 
\noalign{\medskip}{\frac {f}{\omega}}\\ 
\noalign{\medskip}0\\
\noalign{\medskip}0\\
\noalign{\medskip}0\end {array} \right] 
,
\left[ \begin {array}{c} {\frac { \left( k\omega+ifl\right) }{\sqrt {2} \left| k \right| \omega}}\\ 
\noalign{\medskip}{\frac { \left( - ifk+ l\omega \right) }{ \sqrt {2}\left| k \right| \omega}}\\
 \noalign{\medskip}{\frac { \left| k \right| c_+}{\sqrt {2}\omega}}\\
 \noalign{\medskip}0\\ 
 \noalign{\medskip}0\\ 
 \noalign{\medskip}0\end {array} \right] 
,\\
&\left[ \begin {array}{c} 0\\
 \noalign{\medskip}0\\
 \noalign{\medskip}0\\
 \noalign{\medskip}{\frac { \left( -k\omega+ifl \right) }{\sqrt {2} \left| k \right| \omega}}\\
 \noalign{\medskip}{\frac { \left( -ifk-l\omega \right) }{\sqrt {2} \left| k\right| \omega}}\\
 \noalign{\medskip}{\frac { \left| k \right| {c_-}}{\sqrt {2}\omega}}\end {array} \right] 
,
 \left[ \begin {array}{c} 0\\
  \noalign{\medskip}0\\
   \noalign{\medskip}0\\
\noalign{\medskip}-{\frac { il c_-}{\omega}}\\
\noalign{\medskip}{\frac { ik c_-}{\omega}}\\
\noalign{\medskip}{\frac {f}{\omega}}\end {array} \right] 
,
 \left[ \begin {array}{c} 0\\ 
 \noalign{\medskip}0\\ 
 \noalign{\medskip}0\\ 
 \noalign{\medskip}{\frac { \left( k\omega+ ifl \right)}{ \sqrt {2}\left| k \right| \omega}}\\ 
\noalign{\medskip}{\frac { \left( - ifk+ l\omega \right) }{\sqrt {2} \left| k\right|  \omega}}\\ 
 \noalign{\medskip}{\frac { \left| k \right| c_-}{\sqrt {2}\omega}}\end {array} \right] . \addtocounter{equation}{1}\tag{\theequation \textit{a-f}}
\label{2Leigvecs}
\end{align}
Henceforth these modes will be referred to by $\boldsymbol{r}_{\boldsymbol{k}}^{\alpha m}$.

Each vertical mode is equivalent to those in the one layer system (see equation (2.51) in \cite{embid1996averaging}). Properties then transfer from their analysis; the 0 modes are in geostrophic balance, and the fast modes have 0 linear potential vorticity (defined in section \ref{LPV2_ref}). In the remainder of this work we refer to fast modes (inertial gravity waves) and slow modes (potential vorticity modes) for the $\omega\ne0,\omega=0$ modes respectively.

The nonlinear parts for the equations of $\boldsymbol{u}_{ m}$ and $p_{ m}$ are transformed similarly giving:
\begin{align} 
N_u^{ m}&=\sum_{ m_1, m_2}A_{ m_1 m_2}^{ m}(\boldsymbol{u}_{ m_1}\cdot\nabla)\boldsymbol{u}_{ m_2},\label{nlin1} \\
N_p^{ m}&=\sum_{ m_1, m_2}\frac{c_{ m}}{c_{ m_1}}A_{ m_1 m_2}^{ m}\nabla\cdot(p_{ m_1}\boldsymbol{u}_{ m_2}), \label{nlin2}
\end{align}
where 
\begin{align}
A_{ m_1 m_2}^{ m}&=\frac{ m_1 m_2}{ (L_+-L_-)^2}\left[\frac{L_ m}{H_1}+\frac{L_{- m_1}L_{- m_2}}{H_2}\right].
\end{align}
Here there is coupling between all of the different combinations of vertical modes as $A$ is, in general, non-zero.
\subsection{Conservation of potential vorticity}
\label{LPV2_ref}
As remarked previously all coupling between layers in the equations is due to the pressure terms. We define the potential vorticity (hereafter referred to as PV) similarly to that of the one layer equations (see \cite{vallis2006atmospheric}). The pressure term is redundant in calculation of PV and this term is the only part that couples the equations. It follows that there can be no coupling between the PV of each layer; the potential vorticity is conserved per layer:
\begin{align}
\frac{D_iq_i}{Dt}=0, \qquad q_i=\frac{\zeta_i+f}{h_i}, \addtocounter{equation}{1}\tag{\theequation \textit{a,b}}
\end{align}
where ${D_i}/{Dt}$ is the Lagrangian derivative for the flow in the $i$th layer and $\zeta_i=(\nabla\times\boldsymbol{u}_i)_z$ is the relative vorticity. For the linear part of the problem we have:
\begin{align}
\frac{\partial Q_i}{\partial t}=0, \qquad Q_i=\zeta_i-\frac{f\eta_i}{H_i}. \addtocounter{equation}{1}\tag{\theequation \textit{a,b}}
\end{align}
This is identical to the statement of geostrophic balance for two layers.

We now perform the transform (\ref{both_trans}) from layers to the vertical mode basis and find:
\begin{align}
\frac{\partial Q_ m}{\partial t}=0, \qquad Q_ m=\zeta_ m-\frac{fp_ m}{c_{ m}}. \label{lin_pv_mode} \addtocounter{equation}{1}\tag{\theequation \textit{a,b}}
\end{align} \edef\linpvmode{\theequation}

The direct transformation of linear PV is given by:
\begin{align}
Q_ m&=L_{ m}H_1^2Q_1+H_2^2Q_2.
\end{align}
It follows that $Q_1=Q_2=0\iff Q_+=Q_-=0$.

Using (\linpvmode $b$) to calculate the PV associated with the eigenfunctions derived in (\ref{2Leigvecs}) we find that the fast modes have zero linear PV; all PV for the system is contained in the slow modes. This simplifies the calculation of enstrophy in these cases.
\subsection{Integral conservation laws}
In addition to potential vorticity there are two integrally conserved quantities: energy and enstrophy. The quadratic and cubic parts of these are used in section \ref{vanmeth}

In the layer basis enstrophy conservation in flux form is defined as:
\begin{align}
\frac{\partial Z_i}{\partial t}+\nabla\cdot(Z_i \boldsymbol{u}_i)=0, \qquad Z_i=h_iq_i^2. \addtocounter{equation}{1}\tag{\theequation \textit{a,b}}
\end{align}
This can be expanded as follows:
\begin{align}
&Z_i=Z_i^{(2)}+Z_i^{(3)}+...\ ,\\
&Z_i^{(2)}=H_iQ_i^2,\\
&Z_i^{(3)}=-\eta_iQ_i^2. \label{3rd_enstr}
\end{align}

Here if the linear PV in both layers is zero ($Q_i=0$) we see that the enstrophy (up to third order) is also zero. Earlier we observed that this also implies that the linear PV in the mode basis is also zero ($Q_ m=0$). If we combine these two ideas it follows that enstrophy in the mode basis is also zero in this case.

We integrate this across the domain:
\begin{align}
\frac{d \mathcal{Z}_i}{dt}=0, \qquad \mathcal{Z}_i=\int_{D_i} h_iq_i^2 d\boldsymbol{x}.\addtocounter{equation}{1}\tag{\theequation \textit{a,b}}
\end{align}

Energy conservation can be derived from the initial equations by multiplying the momentum equations (\ref{mom1}) and (\ref{mom2}) by $h_i\boldsymbol{u}_i$ and manipulating into the form:
\begin{subequations}
\begin{align}
\frac{\partial E}{\partial t}+\nabla \cdot (\frac{1}{2}(r_\rho h_1|\boldsymbol{u}_1|^2\boldsymbol{u}_1+h_2|\boldsymbol{u}_2|^2\boldsymbol{u}_2+(r_\rho p_1h_1+p_2h_2))=0, \\ 
E=\frac{1}{2}(r_\rho h_1|\boldsymbol{u}_1|^2+h_2|\boldsymbol{u}_2|^2+g(r_\rho\eta_1^2+2r_\rho\eta_1\eta_2+\eta_2^2)),
\end{align}
\end{subequations}
and similarly to enstrophy there is the integral form:
\begin{align}
\frac{d\mathcal{E}}{dt}=0, \qquad \mathcal{E}=\frac{1}{2}\int_D r_\rho h_1|\boldsymbol{u}_1|^2+h_2|\boldsymbol{u}_2|^2+g(r_\rho\eta_1^2+2r_\rho\eta_1\eta_2+\eta_2^2)d\boldsymbol{x}. \addtocounter{equation}{1}\tag{\theequation \textit{a,b}}
\end{align}
\subsection{Rigid lid limit}
\label{rig_lid}
The rigid lid approximation is a specific case of the two layer equations used in geophysical applications. The rigid lid equations are only physically realised when the top layer is at a fixed solid boundary, otherwise they are an approximation based on the (unphysical) assumptions that the gravitational force is large compared to the Coriolis force and the densities of the two layers are close. In this parameter regime the sizes of waves on the external boundary are negligible compared to the size of the waves on the internal layer. The limit $g\rightarrow\infty$ is used to force rigidity in the upper layer when no physical boundary actually exists. This limit is taken separately to the asymptotic limit we are taking; this defines the basic system before any other assumptions are made.  

We start the derivation (see \cite{salmon1998lectures}) by taking $r_\rho\rightarrow1$ and defining the reduced gravity $g'=g(1-r_\rho)$ which we then require to be finite in the limit. The transformation to external and internal modes then becomes:
\begin{align}
L_{ m}&=\begin{cases}1 &  m=+,\\
-\frac{H_2}{H_1} &  m=-,
 \end{cases}\\
 c_{ m}^2&=\begin{cases} g(H_1+H_2)&  m=+,\\
\frac{g\rq{}H_1H_2}{H_1+H_2} &  m=-,
 \end{cases}
\end{align}
by setting $r_\rho=1-\delta$, $g\rq{}=g\delta$, $\delta\ll1$ and taking Taylor series in $\delta$.

Because in this limit we take $g\rightarrow\infty$, the external wave speed becomes infinite (${c_+\rightarrow\infty}$) and the corresponding external Rossby radius of deformation ${\lambda_+={c_+}/{f}\rightarrow\infty}$ as well. By definition, below the Rossby radius of deformation the surface displacement, and hence fast modes, are negligible and so for an infinite radius all fast modes must be neglected.  This corresponds to the external boundary becoming fixed in the longwave limit. Effectively waves cannot propagate on the external boundary and so it becomes \lq{}rigid\rq{}. 

Another consequence of this limit is that the pressure in the external mode becomes undefined, changing the structure of the equations: this pressure can no longer evolve in time. This means that if the external mode is initially defined as motionless the equations only describe the internal mode in a one layer system similar to the usual one.

Neither the one layer nor the rigid lid equations allow for resonant interactions in the fast modes. This is because no resonances can exist between fast mode triads with the same characteristic velocity $c$. Any interactions amongst fast modes will be a new phenomenon in the case of equations with two free layers.
\subsection{Thin layer limit}
Another application relevant to the ocean is to make the upper layer thin compared to the lower layer. However this will change the analysis completely: the thin layer/internal mode will be pushed to the next order of the expansion and so the leading order effects will be equivalent to the one layer case, with corrections at higher order.

To see this consider the amplitude ratio $D/H\sim\epsilon$. If we choose one of our layer depths to be asymptotically small, to avoid violation of this condition we require $D\sim\epsilon^2$ and so all of the dynamics of this layer can only effect the $O(\epsilon^2)$ terms and higher in the non-triad interactions. 

Equivalently setting $H_1=H$, $H_2=H\epsilon$ in (\ref{c_calc}):
\begin{align*}
c_m^2&=gH\begin{cases}1+\frac{r_\rho}{4}\epsilon\ \ \ +O(\epsilon^2)& m=+,\\ (1 -\frac{r_\rho}{4})\epsilon\ +O(\epsilon^2)&m=- ,
\end{cases}\\
L_m&=\begin{cases}1-(\frac{r_\rho}{4}-1)\epsilon\ \ +O(\epsilon^2)& m=+,\\  -\frac{r_\rho}{4}\epsilon\qquad \qquad+O(\epsilon^2)&m=- ,
\end{cases}
\end{align*}
which shows that the internal mode will be asymptotically small and asymptotically slow in our calculations.

\section{Derivation of resonant triads}
\subsection{Multiscale asymptotic expansion}
\label{qg_lim}
Much work already exists analysing the triad resonances of the one layer equations, for instance \cite{warn1986statistical}, \cite{babin1997regularity}, \cite{medvedev1999slow} and \cite{embid1996averaging}. Our extension to two layers recovers similar results, although we will explore the differences in section \ref{fff_sec}. For the usual quasigeostrophic limit it is required that the non-dimensional parameters Rossby number, Froude number, and displacement/depth ratio are all proportional to $\epsilon$ as previously stated in section \ref{bas_eq}. We can then reduce the system to the non-localised form:
\begin{eqnarray}
\frac{\partial \boldsymbol{u}}{\partial t}+\frac{1}{\epsilon}\mathcal{L}(\boldsymbol{u})+\mathcal{N}(\boldsymbol{u},\boldsymbol{u})=0, \label{non_loc3}
\end{eqnarray}
where $\boldsymbol{u}=(u_+,v_+,h_+,u_-,v_-,h_-)^T$ and $\mathcal{L}$ is a linear operator (\ref{mode_lin}) defined above in Fourier space with the set of orthonormal eigenvectors in (\ref{2Leigvecs}). Here the nonlinear terms are represented by the bilinear operator $\mathcal{N}$. This operator has two arguments, but only one input, $\boldsymbol{u}$, and hence we have some freedom in its definition. We choose to define it symmetrically, so that it operates equivalently on the first and second arguments. Using (\ref{nlin1}) and (\ref{nlin2}) (Fourier transformation assumed) we write:
\begin{align}
\mathcal{N}(\boldsymbol{a},\boldsymbol{b})=\frac{1}{2}\left(\begin{array}{c}N_u^{ m}(a,b)+N_u^{ m}(b,a)\\N_p^{ m}(a,b)+N_p^{ m}(b,a)\end{array}\right) \label{nonlinex}
\end{align}
Unlike the usual quasigeostrophic limit, we retain a fast time scale $\tau={t}/{\epsilon}$ as in \cite{embid1996averaging} or \cite{thomas2016resonant} and so write $\boldsymbol{u}=\boldsymbol{u}(\boldsymbol{x},\tau,t)$ as a function of two time scales. Then \ref{non_loc3} becomes:
\begin{eqnarray}
\frac{1}{\epsilon}\left(\frac{\partial \boldsymbol{u}}{\partial \tau}+\mathcal{L}(\boldsymbol{u})\right)=-\left(\frac{\partial \boldsymbol{u}}{\partial t}+\mathcal{N}(\boldsymbol{u},\boldsymbol{u})\right). \label{non_loc}
\end{eqnarray}
As previously mentioned, the different types of eigenfunctions for the operator $\mathcal{L}$ have different behaviours. One set represents fast, gravity waves ($\alpha=\pm$) that are wave-like in fast time $\tau$ with frequency $\omega$. The other set represents slow, PV modes ($\alpha=0$) that do not evolve on the $\tau$ time scale due to the zero eigenvalue. The usual quasigeostrophic equations are derived by taking this limit without using a fast time so that only slow modes are present.

We expand the variable $\boldsymbol{u}$ as follows:
\begin{align}
\boldsymbol{u}(\boldsymbol{x},\tau,t)=\boldsymbol{u}_0(\boldsymbol{x},\tau,t)+\epsilon\boldsymbol{u}_1(\boldsymbol{x},\tau,t)+...\ . \label{sepdef}
\end{align}
Substitution into the non-local equation (\ref{non_loc}) gives the following at each order:
\begin{eqnarray}
O(\epsilon^{-1})\qquad \qquad \qquad \qquad \frac{\partial \boldsymbol{u}_0}{\partial \tau}+\mathcal{L}(\boldsymbol{u}_0)=0, \qquad \qquad \qquad \qquad \ \ \ \\
O(1)\qquad \qquad \frac{\partial \boldsymbol{u}_1}{\partial \tau}+\mathcal{L}(\boldsymbol{u}_1)=-\left(\frac{\partial \boldsymbol{u}_0}{\partial t}+\mathcal{N}(\boldsymbol{u}_0,\boldsymbol{u}_0)\right). \qquad \qquad 
\end{eqnarray}
At first order the equation is linear and we can write the solution in terms of the exponential operator and an average over fast time $\bar{\boldsymbol{u}}$:
\begin{eqnarray}
\boldsymbol{u}_0(\boldsymbol{x},\tau,t)=\bar{\boldsymbol{u}}(\boldsymbol{x},t)e^{-\tau\mathcal{L}},\\
\bar{\boldsymbol{u}}=\lim_{\tau\rightarrow \infty} \int_{0}^{\tau}\boldsymbol{u}_0(\boldsymbol{x},s,t)\ ds.
\end{eqnarray}
Then solving at the next order:
\begin{eqnarray}
\boldsymbol{u}_1e^{\tau\mathcal{L}}=\left. \boldsymbol{u}_1\right|_{\tau=0}-\left(\tau\frac{\partial \bar{\boldsymbol{u}}}{\partial t}+\int_0^\tau \mathcal{N}(\bar{\boldsymbol{u}}e^{-s\mathcal{L}},\bar{\boldsymbol{u}}e^{-s\mathcal{L}})e^{s\mathcal{L}}ds \right).
\end{eqnarray}
With the equation in this form we can identify possible secular terms as any of $O(\tau)$ or higher: those in the round brackets. To maintain the separation of scales for the velocities/pressures as defined in (\ref{sepdef}) these terms must be zero in the limit $\tau \rightarrow \infty$. This is the \lq{}cancellation of oscillations\rq{} concept, used by \cite{schochet1994fast}, where he used the concept to prove convergence for general hyperbolic equations. We will also assume that the vector $\bar{\boldsymbol{u}}$ is written in its eigenbasis such that the matrix exponential is just the exponential of the frequency of the corresponding eigenvalue ($e^{-i\omega t}$):
 \begin{align}
\frac{\partial \bar{\boldsymbol{u}}}{\partial t}&=-\lim_{\tau \rightarrow \infty}\frac{1}{\tau}\int_0^\tau \mathcal{N}(\bar{\boldsymbol{u}}e^{-s\mathcal{L}},\bar{\boldsymbol{u}}e^{-s\mathcal{L}})e^{s\mathcal{L}}ds\\
=&-\lim_{\tau \rightarrow \infty}\frac{1}{\tau}\int_0^\tau \sum_{\substack{ \boldsymbol{k},\boldsymbol{k}_1,\boldsymbol{k}_2\\ \alpha,\alpha_1,\alpha_2\\ \boldsymbol{k}=\boldsymbol{k}_1+\boldsymbol{k}_2}}C_{\boldsymbol{k}_1\boldsymbol{k}_2\boldsymbol{k}}^{\substack{\alpha_1 \alpha_2 \alpha\\  m_1  m_2  m}}\sigma^{\alpha_1  m_1}_{\boldsymbol{k}_1}(t)\sigma^{\alpha_2  m_2}_{\boldsymbol{k}_2}(t)\boldsymbol{r}_{\boldsymbol{k}}^{\alpha  m}e^{i\boldsymbol{k}\cdot\boldsymbol{x}}e^{i(\omega^{\alpha_1  m_1}_{\boldsymbol{k}_1}+\omega^{\alpha_2  m_2}_{\boldsymbol{k}_2}-\omega^{\alpha  m}_{\boldsymbol{k}})\tau}ds, \label{sec_terms}
 \end{align}
 where $\sigma^{\alpha_i  m_i}_{\boldsymbol{k}_i}$ represents the wave amplitude of each eigenfunction and the interaction coefficient is defined as:
 \begin{align} \nonumber
C_{\boldsymbol{k}_1\boldsymbol{k}_2\boldsymbol{k}}^{\substack{\alpha_1 \alpha_2 \alpha\\  m_1  m_2  m}}=\frac{iA_{ m_1 m_2}^{ m}}{2}\Big[&(\boldsymbol{v}_{\boldsymbol{k}_1}^{\alpha_1  m_1}\cdot\boldsymbol{k}_2)(\boldsymbol{v}_{\boldsymbol{k}_2}^{\alpha_2  m_2}\cdot \boldsymbol{v}_{\boldsymbol{k}}^{\alpha  m})+(\boldsymbol{v}_{\boldsymbol{k}_2}^{\alpha_2  m_2}\cdot \boldsymbol{k}_1)(\boldsymbol{v}_{\boldsymbol{k}_1}^{\alpha_1  m_1} \cdot \boldsymbol{v}_{\boldsymbol{k}}^{\alpha  m}) \\+&\frac{c_{ m}}{c_{ m_2}}(\boldsymbol{v}_{\boldsymbol{k}_1}^{\alpha_1  m_1}\cdot(\boldsymbol{k}_1+\boldsymbol{k}_2))p_{\boldsymbol{k}_2}^{\alpha_2  m_2}p_{\boldsymbol{k}}^{\alpha  m}+\frac{c_{ m}}{c_{ m_1}}(\boldsymbol{v}_{\boldsymbol{k}_2}^{\alpha_2  m_2}\cdot (\boldsymbol{k}_1+\boldsymbol{k}_2))p_{\boldsymbol{k}_1}^{\alpha_1  m_1}p_{\boldsymbol{k}}^{\alpha  m}\Big], \label{1lintcoeff}
 \end{align}
 where $\boldsymbol{v}$ is the two-dimensional vector composed of the velocities, and we have expressed (\ref{nonlinex}) explicitly in spectral space. $\alpha_i,\  m_i$ define the modes being considered.
 
The integral simplifies further. In the limit, the integral of all oscillatory contributions exactly cancel to 0 and so the only contributions come from the non-oscillatory constant contributions where:
\begin{align}
\omega^{\alpha_1  m_1}_{\boldsymbol{k}_1}+\omega^{\alpha_2  m_2}_{\boldsymbol{k}_2}-\omega^{\alpha  m}_{\boldsymbol{k}}=0, \label{res_cond}
\end{align}
These are the resonant triads. This leaves the equations:
\begin{align}
\frac{\partial \bar{\boldsymbol{u}}}{\partial t}
=& \sum_{\substack{ \boldsymbol{k},\boldsymbol{k}_1,\boldsymbol{k}_2\\ \alpha,\alpha_1,\alpha_2}}C_{\boldsymbol{k}_1\boldsymbol{k}_2\boldsymbol{k}}^{\substack{\alpha_1 \alpha_2 \alpha\\  m_1  m_2  m}}\sigma^{\alpha_1  m_1}_{\boldsymbol{k}_1}(t)\sigma^{\alpha_2  m_2}_{\boldsymbol{k}_2}(t)\boldsymbol{r}_{\boldsymbol{k}}^{\alpha  m}e^{i\boldsymbol{k}\cdot\boldsymbol{x}}\delta_{\boldsymbol{k}-\boldsymbol{k}_1-\boldsymbol{k}_2}\delta_{\omega-\omega_1-\omega_2}\ ,
 \end{align}
 or in terms of only wave amplitudes:
 \begin{align}
\frac{\partial}{\partial t}\sigma_{\boldsymbol{k}}^{\alpha m}
=& \sum_{\substack{ \boldsymbol{k},\boldsymbol{k}_1,\boldsymbol{k}_2\\ \alpha,\alpha_1,\alpha_2}}C_{\boldsymbol{k}_1\boldsymbol{k}_2\boldsymbol{k}}^{\substack{\alpha_1 \alpha_2 \alpha\\  m_1  m_2  m}}\sigma^{\alpha_1  m_1}_{\boldsymbol{k}_1}(t)\sigma^{\alpha_2  m_2}_{\boldsymbol{k}_2}(t)\delta_{\boldsymbol{k}-\boldsymbol{k}_1-\boldsymbol{k}_2}\delta_{\omega-\omega_1-\omega_2}\ . \label{gen_eqn}
 \end{align}

The only possible resonances are the combinations of modes $(\alpha_1,\alpha_2,\alpha)=$
\begin{align} \nonumber
&\text{a)}\ (0,\pm,\pm),\  (\pm,0,\pm),\\ \label{res_types}
&\text{b)}\ (\pm,\pm,0),\\ \nonumber
&\text{c)}\ (\pm,\pm,\pm),\\ \nonumber
&\text{d)}\ (0,0,0).
\end{align}

The first and second in (\ref{res_types}a) are treated as equivalent due to the symmetry chosen in the interaction coefficient $C$.  Combination (\ref{res_types}b) leads to an interaction term of zero as can be shown by direct substitution of the eigenvectors into C in (\ref{1lintcoeff}). This is done in section \ref{non-lin_coeff}, with an alternate analysis in section \ref{vanmeth} to discern the physical cause of this zero value. 

There are only three types of interactions remaining. Slow-slow-slow (\ref{res_types}d) that define the development of the PV modes over the longer time scale t and fast-slow-fast (\ref{res_types}a) and fast-fast-fast (\ref{res_types}c) that define the scattering of fast modes off a slow mode and interactions amongst themselves. 

For the one layer equations it can be shown (see \cite{warn1986statistical} or \cite{embid1996averaging} for example) that there are no fast-fast-fast resonances, and further, that the equations for the slow part are exactly equivalent to the quasigeostrophic approximation. We show that in the two layer case the slow part again evolves independently of the fast, giving the quasigeostrophic equation (see section \ref{non-lin_coeff}). However unlike the one layer case we show, in section \ref{fff_sec}, that there are interactions amongst the fast waves for our two layer system.

\subsection{Interaction coefficients}
\label{non-lin_coeff}
In this section we give the nonlinear interaction coefficient in (\ref{1lintcoeff}) explicitly for the different possible combinations of modes. This allows us to examine in detail and categorise the different possible nonlinear interactions in this system. The vertical mode parameter $ m_i$ is left general and setting $ m= m_1= m_2$ returns a comparable expression to the one derived in \cite{ward2010scattering} for the one layer case.
For the slow modes:
\begin{subequations}
\begin{align}
C_{\boldsymbol{k}_1\boldsymbol{k}_2\boldsymbol{k}}^{\substack{0,0,0\\  m_1  m_2  m}}
&=\frac{A_{ m_1 m_2}^{ m}c_{ m}(\boldsymbol{k}_2\times\boldsymbol{k}_1)}{2\omega\omega_1\omega_2}\left(c_{ m_1}\frac{\omega_2^2}{c_{ m_2}}-c_{ m_2}\frac{\omega_1^2}{c_{ m_1}}\right)\qquad \qquad \qquad\qquad \qquad \qquad \qquad \\
&=\frac{iA_{ m_1 m_2}^{ m}}{2}\frac{c_{ m}}{\omega}\left(\frac{\omega_2}{c_{ m_2}}(\boldsymbol{v}_1\cdot\boldsymbol{k}_2)+\frac{\omega_1}{c_{ m_1}}(\boldsymbol{v}_2\cdot\boldsymbol{k}_1)\right). \label{slowslowslow}
\end{align}
\end{subequations}
For two fast and a slow mode:
\begin{align} 
C_{\boldsymbol{k}_1\boldsymbol{k}_2\boldsymbol{k}}^{\substack{\alpha_1\alpha_20\\  m_1  m_2  m}}=& \qquad 0, \label{zero_coeff}\\ \nonumber\\ \nonumber
C_{\boldsymbol{k}_1\boldsymbol{k}_2\boldsymbol{k}}^{\substack{\alpha_1 0\alpha\\  m_1  m_2  m}}
=&\frac{iA_{ m_1 m_2}^{ m}}{4c_{ m_2}\omega\omega_1\omega_2|\boldsymbol{k}||\boldsymbol{k}_1|}\big[
(if^2(c_{ m}^2-c_{ m_2}^2)|\boldsymbol{k}|^2(\boldsymbol{k}\times\boldsymbol{k}_1)_z+2if^2c_{ m_2}^2(\boldsymbol{k}\cdot\boldsymbol{k_1})(\boldsymbol{k}\times\boldsymbol{k}_1)_z)\\ \nonumber
&+\alpha_1\omega_1f(c_{ m}^2-c_{ m_2}^2)|\boldsymbol{k}|^2(\boldsymbol{k}_1\cdot\boldsymbol{k})-i\alpha\alpha_1\omega_1\omega c_{ m_2}^2(\boldsymbol{k}_1\times\boldsymbol{k})_z|\boldsymbol{k}_1|^2\\ \nonumber
&+\alpha_1\omega_1fc_{ m_2}^2(\boldsymbol{k}\cdot\boldsymbol{k_1})|\boldsymbol{k}_1|^2+2i\alpha\alpha_1\omega_1\omega c_{ m_2}^2(\boldsymbol{k}_1\times\boldsymbol{k})_z(\boldsymbol{k}_1\cdot\boldsymbol{k})\\ 
&+2\alpha\omega c_{ m_2}^2f (\boldsymbol{k}\times\boldsymbol{k}_1)_z^2 +ic_{ m_2}^2c_{ m}^2|\boldsymbol{k}_1|^2|\boldsymbol{k}|^2(\boldsymbol{k}\times\boldsymbol{k}_1)_z \big] .
\end{align}
Three fast modes:
\begin{align} \nonumber
C_{\boldsymbol{k}_1\boldsymbol{k}_2\boldsymbol{k}}^{\substack{\alpha_1 \alpha_2\alpha\\  m_1  m_2  m}}
&=\frac{iA_{ m_1 m_2}^{ m}}{4\sqrt{2}\omega\omega_1\omega_2|\boldsymbol{k}||\boldsymbol{k}_1||\boldsymbol{k}_2|}\big[\\ \nonumber
&+(\alpha_1\omega_1\alpha_2\alpha\omega\omega_2(|\boldsymbol{k}_1|^2+|\boldsymbol{k}_2|^2)+(\alpha_1\omega_1|\boldsymbol{k}_1|^2+\alpha_2\omega_2|\boldsymbol{k}_2|^2)f^2)(\boldsymbol{k}_1\cdot\boldsymbol{k}_2)\\ \nonumber
&+(2\alpha_1\omega_1\alpha_2\alpha\omega\omega_2+(\alpha_1\omega_1+\alpha_2\omega_2)f^2)(\boldsymbol{k}_1\cdot\boldsymbol{k}_2)^2+f^2(\alpha_2 \omega_2-\alpha_1 \omega_1)(\boldsymbol{k}_1\times\boldsymbol{k}_2)_z^2\\ \nonumber
&+if(\alpha_2\alpha\omega\omega_2+f^2)|\boldsymbol{k}_2|^2(\boldsymbol{k}_2\times\boldsymbol{k}_1)_z
-if(\alpha_1\alpha\omega\omega_1+f^2)|\boldsymbol{k}_1|^2(\boldsymbol{k}_2\times\boldsymbol{k}_1)_z\\ \nonumber
&+c_{ m}^2(\omega_1\alpha_1+\omega_2\alpha_2)|\boldsymbol{k}_2|^2|\boldsymbol{k}|^2|\boldsymbol{k}_1|^2+c_{ m}^2(\omega_1\alpha_1|\boldsymbol{k}_2|^2+\omega_2\alpha_2|\boldsymbol{k}_1|^2)|\boldsymbol{k}|^2(\boldsymbol{k}_1\cdot\boldsymbol{k}_2)\\ 
&+ifc_{ m}^2(|\boldsymbol{k}_2|^2
-|\boldsymbol{k}_1|^2)|\boldsymbol{k}|^2(\boldsymbol{k}_2\times\boldsymbol{k}_1)_z \big].
\label{fff_coeff}
\end{align}

We recover the second version of the slow-slow-slow interactions (\ref{slowslowslow}) from the usual quasigeostrophic equations, as demonstrated in the following. As the fast modes have zero linear PV the restriction to slow modes is equivalent to the assumption that the flow, to first approximation, is solely the geostrophic part. The second part of the quasigeostrophic approximation is to assume the advection of the flow is due only to this geostrophic part (the slow mode interaction), and hence we would expect that the equation for the slow part:
\begin{align}
\frac{\partial \sigma^{0 m}_{\boldsymbol{k}}}{\partial t}+\sum_{1,2}C_{\boldsymbol{k}_1\boldsymbol{k}_2\boldsymbol{k}}^{\substack{0,0,0\\  m_1  m_2  m}}\sigma_{\boldsymbol{k}_1}^{0 m_1}\sigma_{\boldsymbol{k}_2}^{0 m_2}\ \delta_{\boldsymbol{k}-\boldsymbol{k}_1+\boldsymbol{k}_2}=0, \label{320}
\end{align}
is equivalent to the quasigeostrophic equations:
\begin{align}
\frac{ \partial Q_i}{\partial t}+((\boldsymbol{u}_g)_i \cdot \nabla)Q_i=0. \label{321}
\end{align}
To prove this equivalence between (\ref{320}) and (\ref{321}), we start by transforming (\ref{321}) into the mode basis:
\begin{align}
\frac{ \partial Q_{ m}}{\partial t}+\sum_{ m_1 m_2}A_{ m_1 m_2}^{ m}((\boldsymbol{u}_g)_{ m_1} \cdot \nabla)Q_{ m_2}=0.
\end{align}
For a general eigenvector $\boldsymbol{r}_{\boldsymbol{k}}^{0 m}$ we now consider the linear potential vorticity: $Q_{\boldsymbol{k}}^{ m}=-{\omega_{\boldsymbol{k}}^{ m}\sigma_{\boldsymbol{k}}^{0m}}/{c_ m}$ calculated directly from the form in (\ref{2Leigvecs}) using (\linpvmode$b$). With this we then write the quasigeostrophic equation symmetrically in Fourier space to see:
\begin{align}
\frac{\partial \sigma^{0 m}_{\boldsymbol{k}}}{\partial t}&+\sum_{1,2}\frac{iA_{ m_1 m_2}^{ m}}{2}\frac{c_{ m}}{\omega}\left(\frac{\omega_2}{c_{ m_2}}(\boldsymbol{v}_1\cdot\boldsymbol{k}_2)+\frac{\omega_1}{c_{ m_1}}(\boldsymbol{v}_2\cdot\boldsymbol{k}_1)\right)\sigma_{\boldsymbol{k}_1}^{0 m_1}\sigma_{\boldsymbol{k}_2}^{0 m_2}\ \delta_{\boldsymbol{k}-\boldsymbol{k}_1-\boldsymbol{k}_2}\\
&=\frac{\partial \sigma^{0 m}_{\boldsymbol{k}}}{\partial t}+\sum_{1,2}C_{\boldsymbol{k}_1\boldsymbol{k}_2\boldsymbol{k}}^{\substack{0,0,0\\  m_1  m_2  m}}\sigma_{\boldsymbol{k}_1}^{0 m_1}\sigma_{\boldsymbol{k}_2}^{0 m_2}\ \delta_{\boldsymbol{k}-\boldsymbol{k}_1-\boldsymbol{k}_2}=0
\end{align}
This confirms that the two layer quasigeostrophic equations are recovered as they were in the one layer case (see \cite{embid1996averaging}). These are still the usual quasigeostrophic equations in our limit, even though we have included a fast time scale. To give additional insight into why it is possible to separate the equations for the slow and fast evolution, in section \ref{vanmeth} an additional method showing that this must happen is presented, using conservation laws.  

The nonlinear interaction coefficients, calculated in section \ref{fff_sec}, must pair with possible resonances similar to those calculated by \cite{warn1986statistical}. 

Before proceeding, we note that the triad interaction equations (\ref{sec_terms}) can be extended to allow for  near resonances, by replacing the resonance condition (\ref{res_cond}) with:
\begin{align}
\omega^{\alpha  m}_{\boldsymbol{k}}-\omega^{\alpha_1  m_1}_{\boldsymbol{k}_1}-\omega^{\alpha_2  m_2}_{\boldsymbol{k}_2} = \epsilon \Omega_{\boldsymbol{k}_1\boldsymbol{k}_2\boldsymbol{k}}^{\substack{\alpha_1\alpha_2 \alpha\\  m_1  m_2  m}}  
\end{align}
At an exact resonance ($\Omega  =0$) there are secular terms in the expansion proportional to t, and their removal leads to (\ref{sec_terms}). But for  $\Omega \ne 0 $ and  of order unity, a different kind of secular term appears, namely proportional  to $(e^{-i\Omega t} - 1)/\epsilon \Omega $ and these terms also need to be removed. The outcome is
 \begin{align}
\frac{\partial}{\partial t}\sigma_{\boldsymbol{k}}^{\alpha m}
=& \sum_{\substack{ \boldsymbol{k},\boldsymbol{k}_1,\boldsymbol{k}_2\\ \alpha,\alpha_1,\alpha_2}}\frac{-C_{\boldsymbol{k}_1\boldsymbol{k}_2\boldsymbol{k}}^{\substack{\alpha_1 \alpha_2 \alpha\\  m_1  m_2  m}}e^{-i\Omega_{\boldsymbol{k}_1\boldsymbol{k}_2\boldsymbol{k}}^{\substack{\alpha_1\alpha_2 \alpha\\  m_1  m_2  m}}  t}}{i\Omega_{\boldsymbol{k}_1\boldsymbol{k}_2\boldsymbol{k}}^{\substack{\alpha_1\alpha_2 \alpha\\  m_1  m_2  m}} }\sigma^{\alpha_1  m_1}_{\boldsymbol{k}_1}(t)\sigma^{\alpha_2  m_2}_{\boldsymbol{k}_2}(t)\delta_{\boldsymbol{k}-\boldsymbol{k}_1-\boldsymbol{k}_2}\ ,
 \end{align}
 whereas before the summation is over all wavenumbers and modes. At an exact resonance a single triad leads to a periodic exchange of energy between the three wave amplitudes, and the same outcome holds for such a near resonance, see \cite{grimshaw2007nonlinear}. In the remaining sections we shall focus on the exact resonance cases,  but will return to this issue of near resonances in our discussion section.  

\subsection{Resonant triads and conservation laws}
\label{vanmeth}
In this section a different method, based on conservation laws, is used to find the interaction coefficients, with the aim of providing insight into the physical cause of the results of the previous section.

Following the argument in \cite{grimshaw2007nonlinear} (more details in \cite{vanneste1994nonlinear}, and \cite{ripa1981theory}) we define the nonlinear interaction coefficient in terms of the quadratic part of the energy and/or enstrophy. This is done by defining quadratic forms that, as well as giving the first terms of the conservation laws, form an orthogonality condition over the different modes. 

First we consider the linearised problem. The dispersion relation is derived where the different branches give rise to different modes, as in section \ref{vert_bas}. We now write the quadratic part of the energy (or enstrophy) in the form:
\begin{align}
\mathcal{E}^{(2)}=\frac{1}{2}\int_{D}\boldsymbol{u}^\dagger \boldsymbol{E} \boldsymbol{u}\ d\boldsymbol{x},
\end{align}
where $\boldsymbol{u}$ is the velocity, $\boldsymbol{u}^\dagger$ is it\rq{}s conjugate transpose and $\boldsymbol{E}$ is a Hermitian matrix. It can be proven \citep{grimshaw2007nonlinear} that this must obey the following orthogonality relation:
\begin{align}
\boldsymbol{u}_p^\dagger \boldsymbol{E} \boldsymbol{u}_q=E_p\delta_{pq},
\end{align}
where p and q define the mode ($\alpha, m, \boldsymbol{k}$) of the velocities. This relation defines the constants $E_p$.

We now solve the nonlinear problem by taking Fourier transforms and splitting into the eigenmodes from the linear problem. We can use this orthogonality relation to isolate the effect on the amplitude of each mode:
\begin{align}
&\dot{\sigma}_p=\frac{1}{2}\sum_{qr}C_p^{qr}\sigma_q^*\sigma_r^*e^{i\Omega_{pqr}t}\delta_{\boldsymbol{k}_p+\boldsymbol{k}_q+\boldsymbol{k}_r}, \label{van1}\\
&C_p^{qr}=\boldsymbol{u}_p^\dagger \boldsymbol{E}[\boldsymbol{N}(\boldsymbol{u}_q,\boldsymbol{u}_r)+\boldsymbol{N}(\boldsymbol{u}_r,\boldsymbol{u}_q)]^*/E_p,\\
&\Omega_{pqr}=\omega_p+\omega_q+\omega_r, \label{van3}
\end{align}
where $\dot{\sigma}$ denotes differentiation of $\sigma$ with respect to time.

In this analysis there is less emphasis on the exact form of the modes and so we have switched to a more concise notation that absorbs the details of each mode into a simpler form. The $\sigma_a$ are the coefficients of Fourier wave mode $a$ where $a$ contains the information of wavenumber $\boldsymbol{k}$ and mode $(\alpha,  m)$. The $C$ and $\sigma$ are now defined as in the previous sections but in a more compact form: 
\begin{align*}
C_p^{qr}&=C_{\boldsymbol{k}_1\boldsymbol{k}_2\boldsymbol{k}}^{\substack{\alpha_1 \alpha_2\alpha\\  m_1  m_2  m}},\\
p=\{\boldsymbol{k}, \alpha,  m\}, \qquad
q&=\{\boldsymbol{k}_1, \alpha_1,  m_1\}, \qquad
r=\{\boldsymbol{k}_2, \alpha_2,  m_2\}. 
\end{align*}
So far equations (\ref{van1})-(\ref{van3}) are identical to Vanneste\rq{}s work, but also equivalent to \cite{embid1996averaging}; the selection of the orthonormal basis simply chooses the basis in which $\boldsymbol{E}=\boldsymbol{I}$ and $E_p=1$. 

Using the conservation of energy and enstrophy laws: the energy (or enstrophy) can be expanded as:
\begin{align}
\mathcal{E}=\mathcal{E}^{(2)}+\mathcal{E}^{(3)}+...=\frac{1}{2}\sum_p E_p |\sigma_p|^2+\frac{1}{6}\sum_{pqr}S_{pqr}\sigma^*_p\sigma^*_q\sigma^*_re^{i\Omega_{pqr}t}+...\ .
\end{align}
Here the coefficient $S_{pqr}$ is symmetric in it's arguments and can be derived from (\ref{3rd_enstr}) for the enstrophy and similarly for the energy. However, as we shall see later this term will be multiplied by zero at this order of the asymptotics, and so the exact form is not needed for our purpose and we omit it here. We differentiate this in time and, because it is a conserved quantity, this derivative must be equal to zero at each order. The lowest order terms are:
\begin{align}
\dot{\mathcal{E}}^{(2)}&=\frac{1}{2}\sum_{pqr}E_pC_p^{qr}\sigma^*_p\sigma^*_q\sigma^*_re^{i\Omega_{pqr}t} \ \ +\ \  \text{c.c}\ \ +...\ =0, \\
\dot{\mathcal{E}}^{(3)}&=\frac{1}{6}\sum_{pqr} i \Omega_{pqr}S_{pqr}\sigma^*_p\sigma^*_q\sigma^*_re^{i\Omega_{pqr}t} +...=0.
\end{align}
Here we have used equation \ref{van1} to express the time derivative as $\sigma$ terms.

Summing the different permutations over a chosen triad (a, b, c) and then repeating the method for enstrophy:
\begin{align}
E_aC_a^{bc}+E_bC_b^{ac}+E_cC_c^{ba}&=-i\Omega_{abc}S_{abc}, \label{energyconstr}\\
Z_aC_a^{bc}+Z_bC_b^{ac}+Z_cC_c^{ba}&=-i\Omega_{abc}T_{abc}.\label{enstrconstr}
\end{align}
Here $T_{abc}$ is defined equivalently to $S_{abc}$.
We consider the resonant cases $\Omega_{abc}=0$, to the first order time scale. If we initialise our `simulation' with only one triad and we can show that no other triad is excited on our time scale (*) then each  fundamental property (energy, enstrophy) must be conserved per triad.

Fast modes have zero enstrophy contribution: $Z_i=0$. And so in the case of two fast modes (b,c) combining to make a slow mode (a) all that remains in (\ref{enstrconstr}) is: $C_a^{bc}=0$. 

We still need to prove (*). This is an exercise in how resonant triads affect each other. If we start with one resonant triad this is equivalent to showing that another triad with at least one mode in common with the initialised triad does not grow in first order time. All that remains is to consider the two possible cases.

\subsubsection{Case 1: triads with a single node overlap}
\begin{figure}
\centerline{
\includegraphics{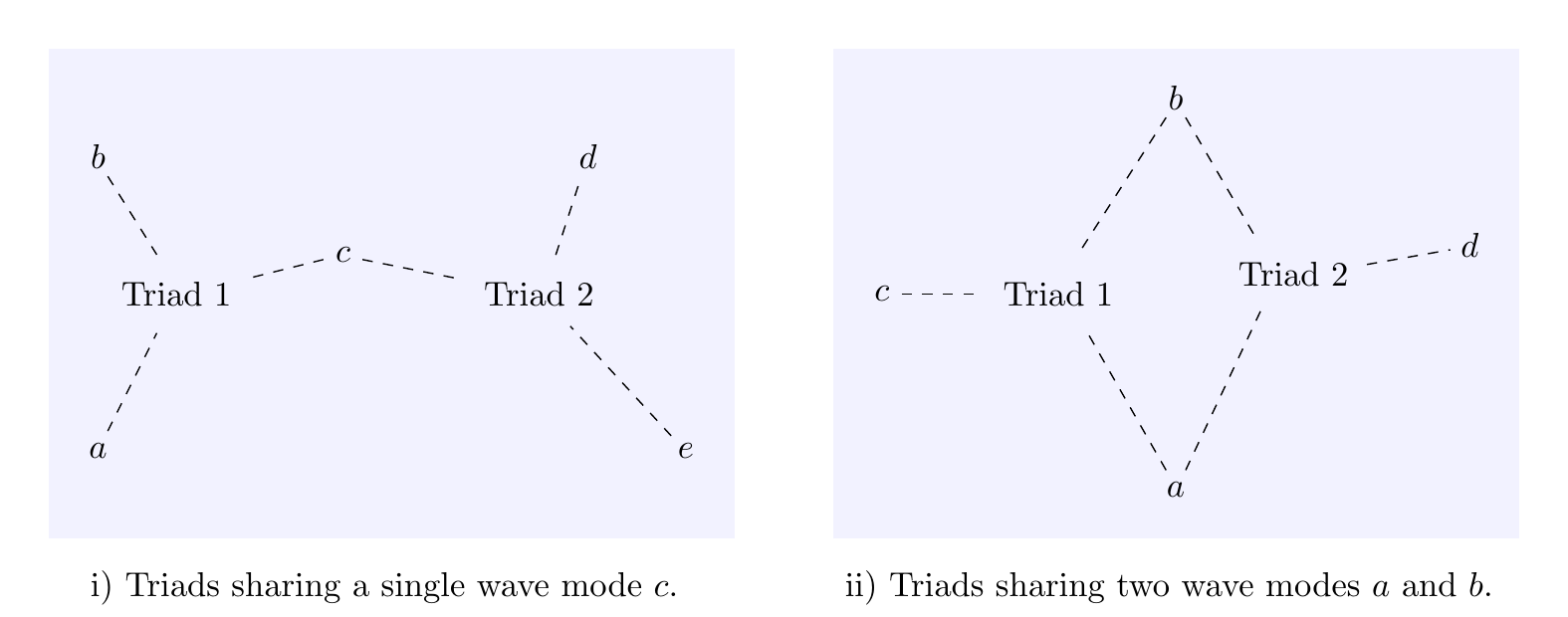}}
\caption{Schematic diagram to show the two options for overlapping triad interactions. i) Case 1: contains five modes with one shared mode between the two triads. ii) Case 2: contains 4 modes with two shared.}
\label{triad_schem}
\end{figure}
In the first case we consider  two triads (with one common mode occurring in both): a total of five modes. We show that if the modes of the second triad are initially zero they must remain zero on that time scale.

The first case is shown schematically in figure \ref{triad_schem}i. We have a single mode (labelled c) belonging to both triads (labelled a, b, c and c, d, e). The set of 5 evolution equations from (\ref{van1}) in terms of wave amplitudes and interaction coefficients are as follows:
\begin{align} \nonumber
\frac{d\sigma_a}{dt}=C_a^{bc}\sigma_b^*\sigma_c^*\ ,&\qquad
\frac{d\sigma_b}{dt}=C_b^{ac}\sigma_a^*\sigma_c^*\ , \qquad
\frac{d\sigma_c}{dt}=C_c^{ba}\sigma_b^*\sigma_a^*+C_c^{ed}\sigma_e^*\sigma_d^*,\\
\frac{d\sigma_d}{dt}&=C_d^{ec}\sigma_e^*\sigma_c^*\ ,\qquad\qquad
\frac{d\sigma_e}{dt}=C_e^{dc}\sigma_d^*\sigma_c^*.\addtocounter{equation}{1}\tag{\theequation \textit{a-e}}
\end{align}
In the case where only the (a, b, c) triad is non-zero initially ($\sigma_d^*=0$, $\sigma_e^*=0$) we have:
\begin{align} \nonumber
\frac{d\sigma_a}{dt}=C_a^{bc}\sigma_b^*\sigma_c^*\ ,& \qquad
\frac{d\sigma_b}{dt}=C_b^{ac}\sigma_a^*\sigma_c^*\ ,\qquad
\frac{d\sigma_c}{dt}=C_c^{ba}\sigma_b^*\sigma_a^*,\\ 
\frac{d\sigma_d}{dt}&=0\ , \qquad\qquad
\frac{d\sigma_e}{dt}=0,\addtocounter{equation}{1}\tag{\theequation \textit{a-e}}
\label{1nodeover}
\end{align}
and so from equations \ref{1nodeover} we see that only the initial triad is evolving on the time scale t. The existence of this type of triad pairing does not affect our ability to isolate a triad.
\subsubsection{Case 2: triads with a double node overlap}
\label{two_ol}
In the second case we consider four modes formed into two triads (two modes must occur in both), and we show that if the mode of the second triad is initially zero then that triad cannot be isolated on this time scale.

Here two nodes (labelled a, b) overlap between triads labelled (a, b, c) and (a, b, d). These are shown schematically in figure \ref{triad_schem}ii. We have a set of 4 evolution equations as follows:
\begin{align} \nonumber
\frac{d\sigma_a}{dt}&=C_a^{bc}\sigma_b^*\sigma_c^*+C_a^{bd}\sigma_b^*\sigma_d^*\ , \qquad
\frac{d\sigma_b}{dt}=C_b^{ac}\sigma_a^*\sigma_c^*+C_a^{ad}\sigma_a^*\sigma_d^*,\\
\frac{d\sigma_c}{dt}&=C_c^{ab}\sigma_b^*\sigma_a^*\ , \qquad\qquad\qquad
\frac{d\sigma_d}{dt}=C_d^{ab}\sigma_a^*\sigma_b^*. \label{4node}\addtocounter{equation}{1}\tag{\theequation \textit{a-d}}
\end{align}

In the case where only the (a, b, c) triad is non-zero initially ($\sigma_d^*=0$) we have:
\begin{align}\nonumber
\frac{d\sigma_a}{dt}&=C_a^{bc}\sigma_b^*\sigma_c^*\ , \qquad
\frac{d\sigma_b}{dt}=C_b^{ac}\sigma_a^*\sigma_c^*,\\
\frac{d\sigma_c}{dt}&=C_c^{ab}\sigma_b^*\sigma_a^*\ , \qquad
\frac{d\sigma_d}{dt}=C_d^{ab}\sigma_a^*\sigma_b^*.\addtocounter{equation}{1}\tag{\theequation \textit{a-d}}
\end{align}

On the time scale $t$ both triads are evolving, and so the four member equation set (\ref{4node}) is needed to describe the motion in $t$. This means we cannot isolate a single triad in this scenario. Therefore we are not able to use the argument (*) from above; a single triad cannot always be considered when triads like these exist, as energy will always pass to other wave numbers on the time scale being considered.
\subsubsection{Application to the two layer equations}
We now apply the theory of the last two subsections to the two layer equations. The equations exhibit this unusual property of non-isolated (non-trivial) resonant triads because there are two branches of the slow mode ($\omega_\pm=0$). We can say that any triad containing a slow mode has a \lq{}sister\rq{} triad containing the other slow mode. By the argument in section \ref{two_ol} we therefore cannot isolate most triads for the two layer analysis. Instead, we isolate the set of nodes that form triads with two overlapping nodes. For fast fast slow interactions this is as above:
\begin{align} \nonumber
\frac{d\sigma_a}{dt}=C_a^{bc}\sigma_b^*\sigma_c^*&+C_a^{bd}\sigma_b^*\sigma_d^*,\qquad
\frac{d\sigma_b}{dt}=C_b^{ac}\sigma_a^*\sigma_c^*+C_a^{ad}\sigma_a^*\sigma_d^*,\\
\frac{d\sigma_c}{dt}&=C_c^{ab}\sigma_b^*\sigma_a^*,\qquad
\frac{d\sigma_d}{dt}=C_d^{ab}\sigma_a^*\sigma_b^*. \addtocounter{equation}{1}\tag{\theequation \textit{a-d}}
\end{align}
Where modes $c$ and $d$ are slow modes with the same wavenumber but different vertical mode $ m$.
 
Performing the energy/enstrophy analysis as above from equations (\ref{energyconstr})\&(\ref{enstrconstr}) then yields:
\begin{align}
(E_aC_a^{bc}+E_bC_b^{ac}+E_cC_c^{ba})\sigma_c^*+(E_aC_a^{bd}+E_bC_b^{ad}+E_dC_d^{ba})\sigma_d^*&=0,\\
(Z_a^+C_a^{bc}+Z_b^+C_b^{ac}+Z_c^+C_c^{ba})\sigma_c^*+(Z_a^+C_a^{bd}+Z_b^+C_b^{ad}+Z_d^+C_d^{ba})\sigma_d^*&=0,\\
(Z_a^-C_a^{bc}+Z_b^-C_b^{ac}+Z_c^-C_c^{ba})\sigma_c^*+(Z_a^-C_a^{bd}+Z_b^-C_b^{ad}+Z_d^-C_d^{ba})\sigma_d^*&=0,
\end{align}
where $Z_i^{ m}$ is the quadratic part of the enstrophy in the $ m$ vertical mode: we have an enstrophy conservation law for each of these.

Using $Z_a^\pm=0$, $Z_b^\pm=0$ (fast modes have zero enstrophy) we have:
\begin{align}
Z_c^{ m_c}C_c^{ba}\sigma_c^*+Z_d^{ m_c}C_d^{ba}\sigma_d^*=0,\\
Z_c^{ m_d}C_c^{ba}\sigma_c^*+Z_d^{ m_d}C_d^{ba}\sigma_d^*=0,
\end{align}
then we solve this linear system recalling that $Z_c^{ m_i}\ne0$ and $Z_d^{ m_i}\ne0$ and ${Z_c^{ m_i}Z_d^{- m_i}-Z_c^{- m_i}Z_d^{ m_i}\ne0}$. It follows that the interaction coefficients here must then both be zero. This implies that the fast-fast-slow interaction coefficient value of zero is a direct result of conservation of quadratic enstrophy.

\subsection{The fast-fast-fast interactions}
\label{fff_sec}
\begin{figure}
\centerline{
\includegraphics{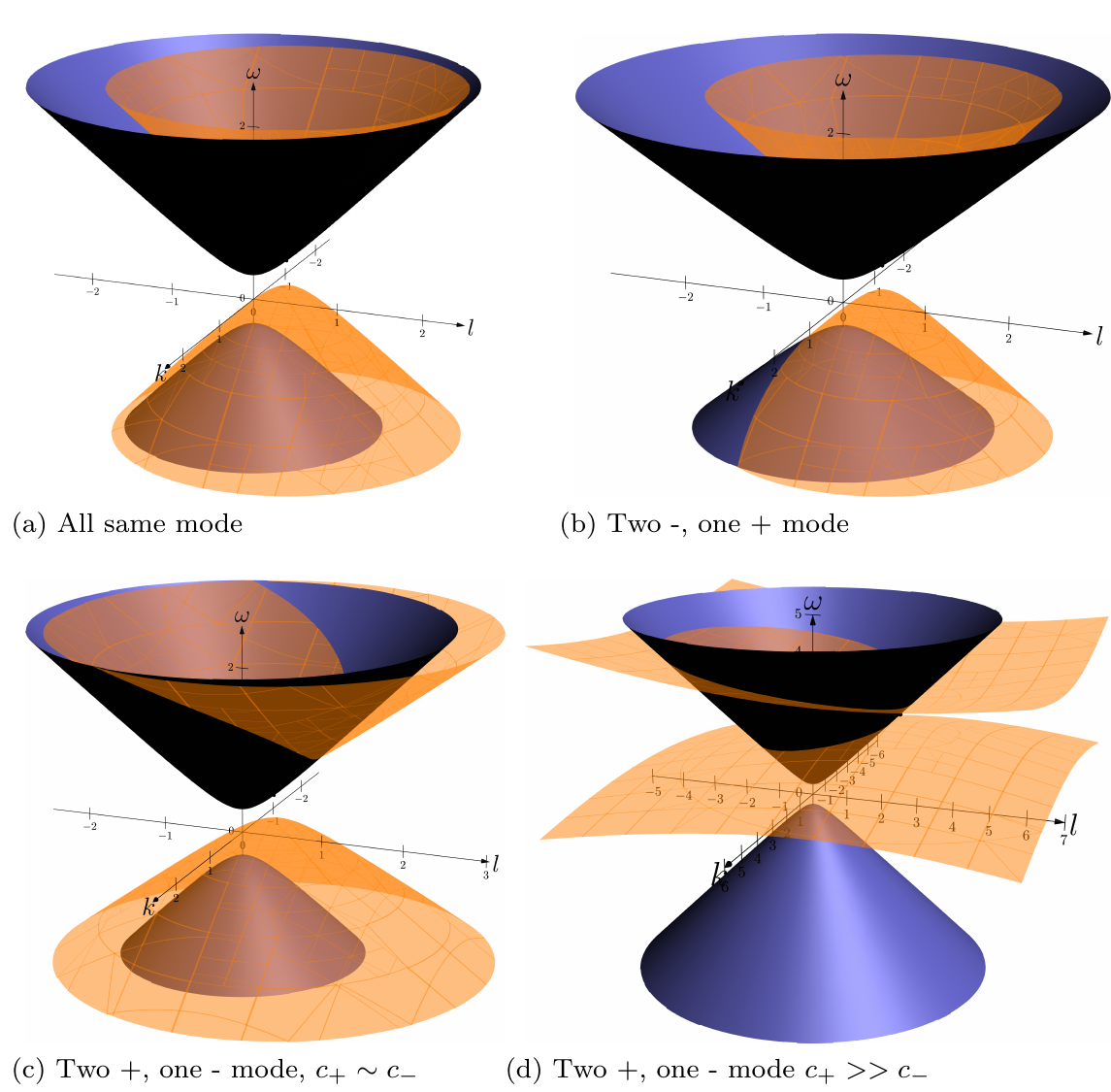}}
\caption{Graphical method to find possible resonant triads in the fast-fast-fast interactions. The dark hyperboloid is the manifold on which the $\omega_1$ may lie (defined by the dispersion relation for $\boldsymbol{k}_1$), the light hyperboloid is centred around a chosen $\omega_1$ in the first manifold. The light hyperboloid therefore represents all possible solutions $\omega\ (=\omega_1+\omega_2)$ and any intercept with the dark hyperboloid represents a possible resonance where $\omega$ has the same vertical mode type as $\omega_1$.}
\label{cones}
\end{figure}

We now return to a key point raised at the end of section \ref{non-lin_coeff}: the fast-fast-fast resonances. These are a clear difference to the single layer version of the shallow water equations, where they cannot occur. We now consider where these resonances are permitted by the dispersion relations. The fast-fast-fast resonances were originally considered by \cite{ball1964energy} for the simpler case with no Coriolis force. Figure \ref{cones} shows graphically how these resonances can exist (this graphical method was discovered independently by researchers in different fields, for example \cite{ziman1960electrons} and \cite{ball1964energy}). In addition to the graphical method, the full derivation of the resonances is shown in Appendix \ref{fffapp}. Case a) in the diagram shows that there are no resonances between waves of the same vertical mode, this is equivalent to the one layer case.

We find that the resonances always exist for any combination of fast modes with different vertical modes: cases b), c), and d) in figure \ref{cones}. In addition there is another more unusual resonance (seen in the lower sheet of the light hyperboloid in case (d) in figure \ref{cones}) where the ratio of the input and output wave speeds is less than 1 (${c_{ m_2}}/{c_{ m}}<1$). Where the output wave is an external wave and one of the inputs is internal if we then consider sufficiently large values of the Burger number for the external mode (${Bu={c_{+}^2|\boldsymbol{k}_1|^2}/{f^2}={L_r^2}/{L^2}}$) this resonance will exist. This condition corresponds to wavelengths at least $\sqrt{3}$ times smaller than the radius of deformation (see appendix \ref{fffapp}). 

Alternatively the equivalent resonance also exists where the two input waves are of different type to the output. Here the ratio of input to output wave speeds needs to be greater than 1 and so the input waves are both external modes. We then require that the wavelength of one input mode is such that the external Burger number is sufficiently large: $Bu={c_{+}^2|\boldsymbol{k}_1|^2}/{f^2}$. 

\begin{figure}
\centerline{
\includegraphics{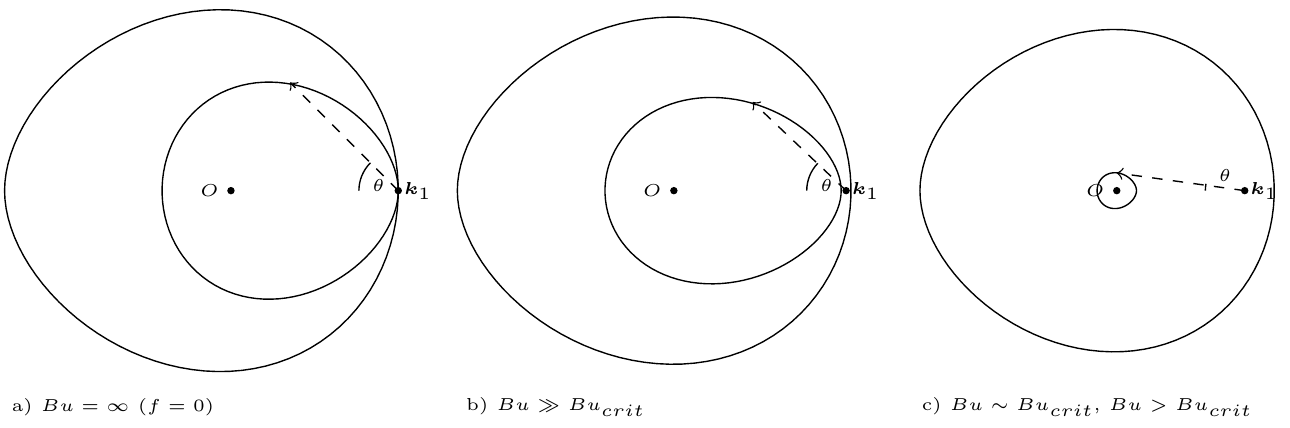}}
\caption{Projection onto wave space of the intersections of the surfaces in figure \ref{cones}d showing the two sets of resonances. a) The non-rotating case equivalent to the diagram in \cite{ball1964energy}, b) The case for large external Burgers number, and c) shows Burger number close to the critical value of the Burgers number such that the angle of incidence in the resonance must be small.}
\label{x_sec}
\end{figure}

These resonances are unusual in that they only exist for angles of incidence within a range around $\pm\pi$. Figure \ref{x_sec} shows the intersections of the surfaces from figure \ref{cones}d  projected into the ($k$, $l$) plane and shows more clearly the angle dependence of the resonance. It can be seen that without the Coriolis force the angle of incidence is in a range ($-\pi/2$,$\pi/2$) but as the Coriolis force becomes more dominant the range of angles is limited to be closer to $-\pi$.

\begin{figure}
\centerline{
\includegraphics[width=\textwidth]{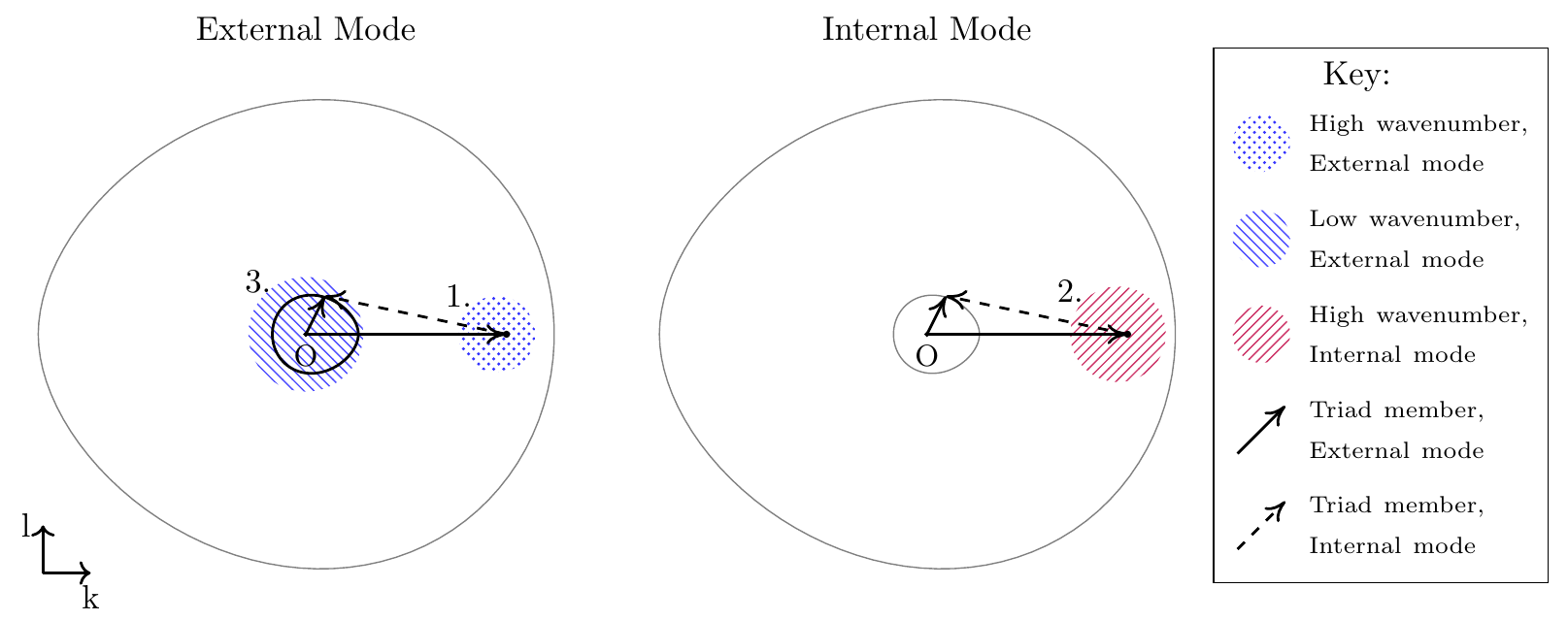}}
\caption{Diagram to show heuristically an example of the directional resonance discussed in section \ref{fff_sec}, on the traces shown in figure 4c. Consider initially that there is a higher proportion of wave energy in external waves of high wavenumber at 1. This can interact via the directional resonance to stimulate a region in the internal modes at 2. Regions 1 \& 2 also resonate and so the waves in region 3 are augmented. The resonances in all three areas stimulate each other and a mechanism similar to Hasselmann\rq{}s criterion will control the dynamics between these regions until other resonances (such as the resonance shown in grey) spread the wave energy out to other areas of wave space.}
\label{blob}
\end{figure}

Figure \ref{blob} shows a possible mechanism based on an input of fast modes at a high wavenumber in the external mode with background modes at all wavenumbers. Via the directional resonance with low wavenumber external waves this would excite internal fast modes at a similar wavenumber to the initial input. These two high wavenumber modes are then able to interact with greater strength to stimulate the low wavenumber external modes, reinforcing the mechanism. These three regions would interact similarly to an exact triad with energy passed amongst themselves, with the lower amplitude parts gaining energy in a similar mechanism to that in Hasselmann\rq{}s criterion for wave growth \citep{hasselmann1967criterion}.

The full equations describing the first closure are:
\begin{align}
\frac{\partial \sigma^{0 m}_{\boldsymbol{k}}}{\partial t}+&\sum_{1,2}C_{\boldsymbol{k}_1\boldsymbol{k}_2\boldsymbol{k}}^{\substack{0,0,0\\  m_1  m_2  m}}\sigma_{\boldsymbol{k}_1}^{0 m_1}\sigma_{\boldsymbol{k}_2}^{0 m_2}\ \delta_{\boldsymbol{k}-\boldsymbol{k}_1-\boldsymbol{k}_2}=0, \label{full1}\\ \nonumber
\frac{\partial \sigma^{\alpha m}_{\boldsymbol{k}}}{\partial t}+\sum_{1,2}C_{\boldsymbol{k}_1\boldsymbol{k}_2\boldsymbol{k}}^{\substack{0,\alpha_2,\alpha\\  m_1  m_2  m}}&\sigma_{\boldsymbol{k}_1}^{0 m_1}\sigma_{\boldsymbol{k}_2}^{\alpha_2  m_2}\ \delta_{\boldsymbol{k}-\boldsymbol{k}_1-\boldsymbol{k}_2}\delta_{\omega-\omega_2}\\
&+\sum_{1,2}C_{\boldsymbol{k}_1\boldsymbol{k}_2\boldsymbol{k}}^{\substack{\alpha_1,\alpha_2,\alpha\\  m_1  m_2  m}}\sigma_{\boldsymbol{k}_1}^{\alpha_1 m_1}\sigma_{\boldsymbol{k}_2}^{\alpha_2  m_2}\ \delta_{\boldsymbol{k}-\boldsymbol{k}_1-\boldsymbol{k}_2}\delta_{\omega-\omega_1-\omega_2}=0, \label{full2}
\end{align}
with the interaction coefficients from section \ref{non-lin_coeff}. 

In order to examine the strength of the new resonance within the full equation 3.49 we position the fast-fast-fast resonance against the fast-slow-fast resonance by numerical evaluation of the size of the interaction coefficient in the case of each triad for the same given wavelengths. The parameters were chosen to be applicable in an oceanic context. In table 1 the maximum absolute value of the different interaction coefficients in any permutation of mixed vertical modes is given. Assuming that all mode amplitudes are within an order of magnitude of each other, this should scale like the change in time of each part of the reduced equations. We find that both interactions have a similar order of magnitude. However it should be noted that if we choose a single vertical mode for all of the constituent modes a larger interaction coefficient can occur for the catalytic case. This suggests that although the overall dynamics may be dominated by the same resonances present in the one layer case, the new resonance is important in evaluating energy exchange between the baroclinic and barotropic modes.

\begin{table}
\begin{tabular}{|c|c|c|}
Wavenumbers and vertical mode & Max $|C^{\pm\pm\pm}|$&  Max $|C^{\pm 0 \pm}|$\\
\hline
$(500,0,+)$, $(-15.21,10,-)$, $(484.79,10,+)$ & $8.04\times10^{-10}$  & $ 3.02\times10^{-10}$\\
\\
$(500,0,+)$, $(-15.21,10,+)$, $(484.79,10,+)$ & - &  $ 3.05\times10^{-7}$\\ 
\\
$(500,0,+)$, $(-122.49,300,-)$, $(377.51,300,+)$ & $6.07\times10^{-10}$ &  $ 3.99\times10^{-10}$\\
\\
$(500,0,+)$, $(-122.49,300,+)$, $(377.51,300,+)$ & - &  $ 4.03\times10^{-7}$
\end{tabular}
\caption{Size of the interaction coefficients for the given wavenumbers, chosen to form a resonant triad of fast-fast-fast modes and a near-resonant triad for sets of fast-slow-fast and modes. Physical parameters used were as follows: $g=10ms^{-2}$, $f=0.0001s^{-1}$, $H_1=500m$, $H_2=4000m$, $L=100km$, the non-dimensional wavenumbers are quoted as $\boldsymbol{k}$ where the physical wavenumber is $2\pi \boldsymbol{k}/L$}
\label{int_coeff_tab}
\end{table}

\section{Discussion}
In this paper, we have examined triad resonances in a rotating shallow water two layer fluid model. The explicit forms of the nonlinear interaction coefficients were found for the different combinations of modes.

Triads were found for combinations of three fast inertial gravity modes, in contrast to the one layer system. These resonances are equivalent to those found by \cite{ball1964energy} in the non-rotating case. Unlike those found by Ball, for certain parameter regimes some of these triads showed unusual behaviour with waves interacting preferentially with waves of a small angle of incidence. In addition this resonance ceases to exist for wavelengths more than some factor greater than the Rossby deformation radius. This resonance is very likely to always be present in geophysical applications with large deformation radii. However in other planets where the length scales and planetary rotation rates may be different, it could be possible to have parameter regimes such that certain resonances exist for different wavenumbers at different latitudes. Particularly interesting are the cases with parameters such that this resonance affects very small wavenumbers. In these cases the resonance is almost entirely between waves with angles of incidence close to zero. In addition this resonance can be arbitrarily strong, dependent on the size of the other waves, due to the $|\boldsymbol{k}|$, $|\boldsymbol{k}_2|$ terms in the interaction coefficient (\ref{fff_coeff}). This describes a possible mechanism for energy to be transferred to low wavenumbers.

The directionality of this new resonance is an unusual feature. The system is intrinsically symmetric, being in an f plane, it is an interesting conclusion that it could be anisotropic, particularly as the mechanism described heuristically in section \ref{fff_sec} would seem to provide a positive feedback mechanism onto itself. 

However, similarly to the one layer case the slow modes were shown to behave exactly as the quasigeostrophic equations for two layers. The weakly nonlinear approximation allows the motion to be split into two equations, the quasigeostrophic equation for the slow part and equations describing the interactions of the fast modes with the slow modes and amongst themselves.

Work on stratified flow using the Boussinesq approximation exists in \cite{mccomas1977resonant} where a similar triad was investigated. However in that and related subsequent work by \cite{bartello1995geostrophic} amongst others, the inclusion of vertical wavenumbers means that the form of the triads are fundamentally different from the present case. Nonetheless, interaction between gravity waves of different vertical modes mimics transfer of gravity wave energy vertically, and is an equivalent process in the present case. This also correlates with the observation that in the continuously stratified equations there are no interactions between gravity waves lying in the same horizontal plane (\cite{lelong1991internal}).This suggests that the asymptotically expanded two layer equations can act as a proxy for understanding of the fully stratified case.

Our \lq{}critical Burger number\rq{} condition can be interpreted as a maximum wavenumber at which this particular resonance will exist. However we can also show that a more general, minimum wavenumber at which fast fast fast resonances occur must exist. For this we simply take the limit $|\boldsymbol{k}| \rightarrow 0$ in the resonance condition:
\begin{align*}
\omega &= f+\frac{c_{m}^2}{2f}|\boldsymbol{k}|^2+... \sim f
\end{align*} 
and this clearly shows that no resonances can exist, provided the Burger number of each wave involved is sufficiently small to make this approximation. In most applications we might expect that this limit will not be reached due to the large size of the Burger number, although it might arise for large fast-rotating planets.

A fundamental issue arises when considering numerical simulations of the equations (\ref{full1}-\ref{full2}), using discrete wavenumbers. The dispersion relation is a function of $\boldsymbol{k}$ which can take all real values. But when we have discrete values of $\boldsymbol{k}$ such as in a numerical simulation of the equations, or when there is a periodic domain, we have a countable set of frequencies $\omega$ in our model, forming a countable subset of the reals. However there is no guarantee that the corresponding frequencies will be resonant and in general only near resonances may appear; indeed we find that even if we choose parameter values to ensure that some particular triad is exactly resonant, it does not automatically follow that any other exact resonant triad appears in the discrete set of wavenumbers. The entire resonance may be absent, removing its physical effect from the model. This observation has been previously made in \cite{smith1999transfer}. Investigations of this nature into the existence of resonant sets in discrete domains have been carried out, details of which can be found in \cite{kartashova2010nonlinear} for example. However the progress made with this number theoretic issue is mostly restricted to problems in which the dispersion relation is proportional to a rational power of $\boldsymbol{k}$, however this is not the case here, and we cannot establish the existence of resonant sets in this way. This is an interesting  problem that seems to be due to the interplay between resonances and a periodic domain. This is a good example of the limits of exact resonances compared to near-resonances: with any exactly resonant theory it may not be possible to simulate on a periodic domain without losing physical effects. Near-resonant interactions (as in for example \cite{smith2005near}) would reintroduce missing resonances, and hence missing physics into the simulations. 

 Further work could investigate the similar system but considering a wave packet to allow continuous wave numbers in the multiple scales analysis. This would introduce the group velocity and the resonance might not occur as wave packets could propagate to different domains without interacting, similarly to the quartet resonance in \cite{zeitlin2003nonlinear}. Comparison to a continuously stratified fluid may also be illuminating for this subject.

An additional curiosity of the equations comes from the energy and enstrophy analysis of section \ref{vanmeth}. As seen in (\ref{3rd_enstr}) the higher orders of enstrophy all contain the linear PV term twice. In the conservation of enstrophy (\ref{enstrconstr}) if we substitute in for two fast modes then the higher order part $T_{abc}$ is identically zero. This means that the resonant triad assumption ($\Omega_{abc}=0$) for this case (calculated explicitly in (\ref{zero_coeff})) is unnecessary to show that the nonlinear interaction coefficient is zero. This may have additional consequences when the analysis is extended to higher orders: the form of the conserved enstrophy terms implies triads preserve enstrophy to higher than quadratic order.

At higher orders of the expansion one might expect that terms will appear in which the fast wave modes influence the slow, as found in the one layer system by \cite{thomas2016resonant}. If we consider only the set of modes in one vertical mode the dynamics behaves identically to a single layer. Because of this, in terms of wave interactions, we don\rq{}t expect to lose behaviours within the two layer model: all the dynamics from the one layer model will be present plus any additional interactions.

\appendix
\section{}\label{appA} 
\subsection{Fast-fast-fast resonances}
\label{fffapp}
In the one layer case this combination can be shown to be impossible (see \cite{warn1986statistical} or \cite{ward2010scattering} for example). However in the two layer case the different wave speeds allow this possibility. We seek to solve:
\begin{align}
\alpha\omega&=\alpha_1\omega_1+\alpha_2\omega_2, \label{resonances}
\end{align}
where $\alpha_i=\pm1$.
We substitute the relevant branches of the dispersion relation:
\begin{align}
\alpha\sqrt{c_{ m}^2|\boldsymbol{k}|^2+f^2}&=\alpha_1\sqrt{c_{ m_1}^2|\boldsymbol{k}_1|^2+f^2}+\alpha_2\sqrt{c_{ m_2}^2|\boldsymbol{k}_2|^2+f^2},
\end{align} 
we square both sides and rearrange:
\begin{align}  
c_{ m}^2|\boldsymbol{k}|^2-c_{ m_1}^2|\boldsymbol{k}_1|^2&-c_{ m_2}^2|\boldsymbol{k}_2|^2-f^2=2\alpha_1\alpha_2\sqrt{c_{ m_1}^2|\boldsymbol{k}_1|^2+f^2}\sqrt{c_{ m_2}^2|\boldsymbol{k}_2|^2+f^2},
\end{align} 
we square again:
\begin{align} 
(c_{ m}^2(|\boldsymbol{k}_1|^2+2(\boldsymbol{k}_1\cdot\boldsymbol{k}_2)&+|\boldsymbol{k}_2|^2)-c_{ m_1}^2|\boldsymbol{k}_1|^2-c_{ m_2}^2|\boldsymbol{k}_2|^2-f^2)^2=4(c_{ m_1}^2|\boldsymbol{k}_1|^2+f^2)(c_{ m_2}^2|\boldsymbol{k}_2|^2+f^2),
\end{align} 
we expand and gather terms:
\begin{align} \nonumber
(c_{ m}^2-c_{ m_1}^2)^2&|\boldsymbol{k}_1|^4+(c_{ m}^2-c_{ m_2}^2)^2|\boldsymbol{k}_2|^4\\ \nonumber
+4c_{ m}^2(c_{ m}^2-c_{ m_1}^2)&(\boldsymbol{k}_1\cdot\boldsymbol{k}_2)|\boldsymbol{k}_1|^2+4c_{ m}^2(c_{ m}^2-c_{ m_2}^2)(\boldsymbol{k}_1\cdot\boldsymbol{k}_2)|\boldsymbol{k}_2|^2\\ \nonumber
+4c_{ m}^4(\boldsymbol{k}_1\cdot\boldsymbol{k}_2)^2&+2(-c_{ m_2}^2c_{ m_1}^2+c_{ m}^4-c_{ m}^2c_{ m_1}^2-c_{ m_2}^2c_{ m}^2)|\boldsymbol{k}_1|^2|\boldsymbol{k}_2|^2\\
-2(c_{ m}^2+c_{ m_1}^2)|\boldsymbol{k}_1|^2f^2&-2(c_{ m}^2+c_{ m_2}^2)|\boldsymbol{k}_2|^2f^2
-4c_{ m}^2(\boldsymbol{k}_1\cdot\boldsymbol{k}_2)f^2-3f^4=0.
\end{align} 
We define $K={|\boldsymbol{k}_2|}/{|\boldsymbol{k}_1|}$, $R_1={c_{ m_1}^2}/{c_{ m}^2}$, $R_2={c_{ m_2}^2}/{c_{ m}^2}$ and $F={f}/{c_{ m}|\boldsymbol{k}_1|}$ with $\theta$ as the angle between the two input wave vectors. Writing the equation as a quartic in K:
\begin{align} \nonumber
& \ \ \ \ \ \ \ \ AK^4+BK^3+CK^2+DK+E=0,\\ \nonumber
A&=(1-R_2)^2 ,\\ \nonumber
B&=4(1-R_2)\cos{\theta},\\
C&=4\cos^2\theta+2(1-R_2)(1-R_1)-4R_1R_2-2(1+R_2)F^2, \label{quart} \\ \nonumber
D&=4((1-R_1)-F^2)\cos\theta,\\ \nonumber
E&=(1-R_1)^2-2(1+R_1)F^2-3F^4.
\end{align}
There are three distinct cases to consider:
\begin{enumerate}
\item $R_1=R_2=1$ (Reduces to the one layer case - no solution)
\item $R_1=1\ne R_2$ 
\item $R_1=R_2\ne 1$
\end{enumerate}
Within each of these cases $R_i>1$ $R_i<1$ need to be considered.\\

\begin{figure}
\centerline{
\includegraphics{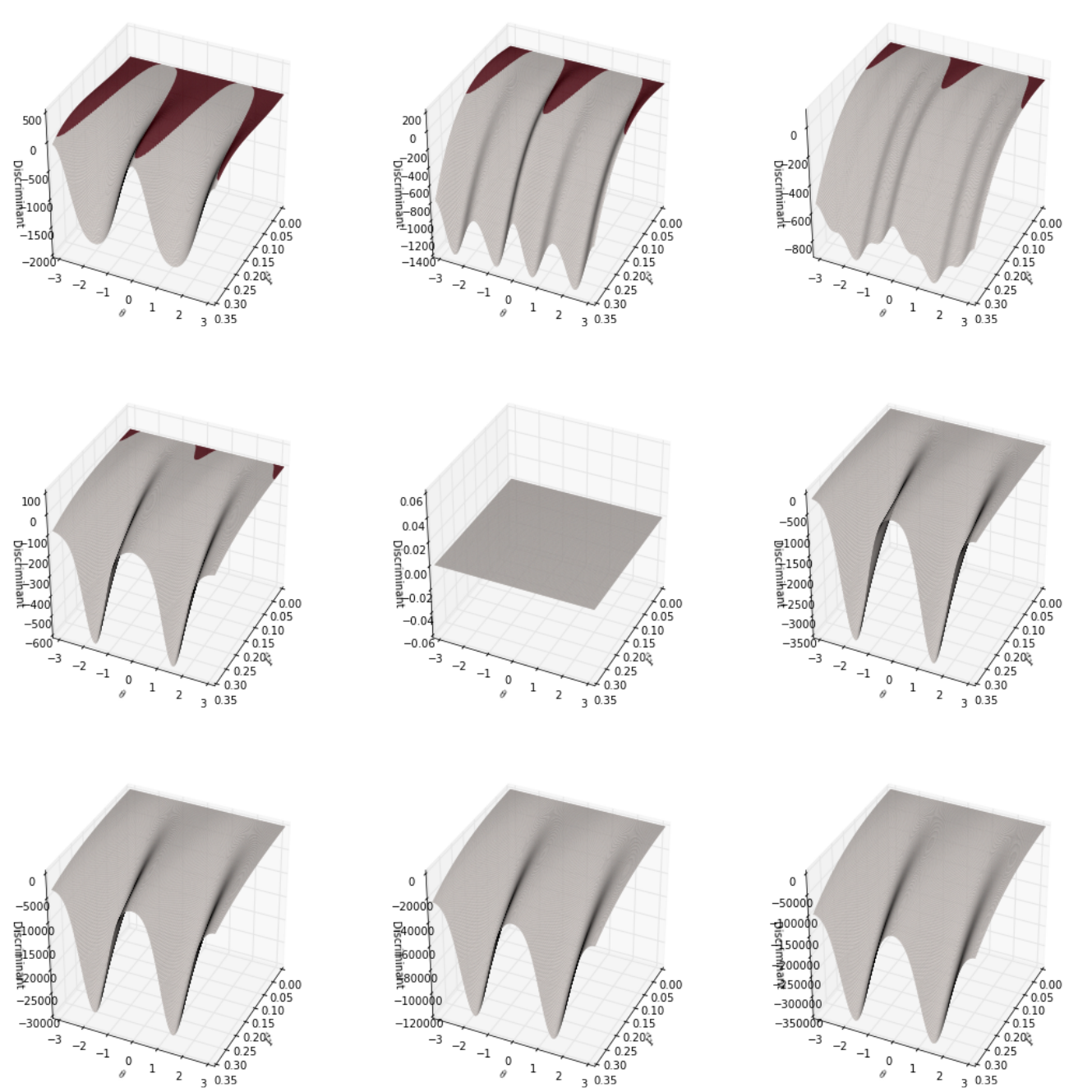}}
\caption{$\Delta$ for $R_2=\{0,0.25,0.5,0.75,1,1.25,1.5,1.75,2\}$ represented as surfaces in parameter space $\{F^2,\theta\}$. Black (red in coloured online version) shows values greater than 1.}
\label{Delta_diag}
\end{figure}

Considering the second case $R_1=1$ Equations \ref{quart} become:
\begin{align} \nonumber
& \ \ \ \ \ \ \ \ AK^4+BK^3+CK^2+DK+E=0,\\ \nonumber
A&=(1-R_2)^2, \\ \nonumber
B&=4(1-R_2)\cos{\theta},\\
C&=4\cos^2\theta-4R_2-4F^2,\\ \nonumber
D&=-4F^2\cos\theta,\\ \nonumber
E&=-4F^2-3F^4. 
\end{align}
We define q, s as the coefficients of the reduced quartic as found in \cite{rees1922graphical} then for all parameter values our equation has: $q<0$, $p=s-{q^2}/{4}<0$. This means that there are either 2 or 4 real solutions when the quartic discriminant $\Delta$ is less than or greater than 0 respectively. This is plotted in figure \ref{Delta_diag}. The areas of 4 solutions correspond to the second crossing point shown in the conic diagrams in figure \ref{cones}. The largest value of $F^2$ with 4 solutions occurs at $R_2=0$, $\cos^2\theta=1$ where $F^2=1/3$.

The definitions of $\Delta$, $q$ and $p$ are:
\begin{align}
q&=8AB-3B^2,\\
p&=64A^3E-16A^2C^2+16AB^2C-16A^2BD-3B^4,\\ \nonumber
\Delta&=256A^3E^3-192A^2BDE^2-128A^2C^2E^2+144A^2CD^2E\\ \nonumber
            &-27A^2D^4+144AB^2CE^2-6AB^2D^2E-80ABC^2DE\\ \nonumber
            &+18ABCD^3+16AC^4E-4AC^3D^2-27B^4E^2+18B^3CDE\\ 
            &-4B^3D^3-4B^2C^3E+B^2C^2D^2.
\end{align}
In full:
\begin{align}
q=&-16(R_2 - 1)^2(F^2(R_2 + 1) + 2R_2 + \cos^2\theta)\le0, \\
p=&-256(F^2 + 1)(R_2 - 1)^4(F^2(R_2^2 - R_2 +1) + R_2^2 + 2R_2\cos^2\theta)\le0,\\ \nonumber
\Delta=& -4096F^2(1+F^2)^2(R_2-1)^2\Big[4R_2(R_2-\cos^2\theta)^3\\ \nonumber
&+(R_2^2(11R_2^2-8R_2+8)-2R_2(R_2+10)(2R_2-1)\cos^2\theta+(2R_2-1)(10R_2+1)\cos^4\theta)F^2\\ \nonumber
&+(2(R_2^2-R_2+1)(5R_2^2-2R_2+2)+2(2R_2-1)(R_2^2-7R_2+1)\cos^2\theta)F^4\\ 
&+3(R_2^2-R_2+1)^2F^6\Big].
\end{align}
So with reference to the original resonance equation (\ref{resonances}) to recover $\alpha$ values we find that there are three distinct types of resonances:
\begin{enumerate}
\item{$R_2>1$: where $\alpha=\alpha_2=-\alpha_1$}
\item{$R_2<1$: where $\alpha=\alpha_2=\alpha_1$}
\item{$R_2<1$, $\Delta>0$: where $\alpha=\alpha_1=-\alpha_2$}
\end{enumerate}
It is then possible to map these resonances onto those with $R_1=R_2\ne1$ by rearranging the input and output waves and relabelling:
\begin{align}
&\alpha_1\omega_1(\boldsymbol{k}_1)+\alpha_2\omega_2(\boldsymbol{k}_2)=\alpha\omega(\boldsymbol{k}),\\
\implies&\alpha_1\omega_1(\boldsymbol{k}_1)-\alpha\omega(\boldsymbol{k})=-\alpha_2\omega_2(\boldsymbol{k}_2),\\
\implies&\alpha_a\omega_a(\boldsymbol{k}_a)+\alpha_b\omega_b(\boldsymbol{k}_b)=\alpha_c\omega_c(\boldsymbol{k}_c),
\end{align}
where $\alpha_a=\alpha_1$, $\alpha_b=-\alpha$, $\alpha_c=-\alpha_2$ and the wavenumber vectors are ${\boldsymbol{k}_a=-\boldsymbol{k}_1}$, ${\boldsymbol{k}_b=\boldsymbol{k}=\boldsymbol{k}_1+\boldsymbol{k}_2}$, $\boldsymbol{k}_c=\boldsymbol{k}_2$. In the new set of resonances ${R_a={c_{ m_1}^2}/{c_{ m_2}^2}}$, ${R_b={c_{ m}^2}/{c_{ m_2}^2}\left(=R_2^{-1}\right)}$, ${F={f}/{c_{ m}|\boldsymbol{k}_1|}}$.
We then have the equivalent set of 3 resonances:
\begin{enumerate}
\item{$R_a=R_b<1$: where $\alpha_b=\alpha_c=\alpha_a$}
\item{$R_a=R_b>1$: where $\alpha_b=\alpha_c=-\alpha_a$}
\item{$R_a=R_b>1$, $\Delta>0$: where $-\alpha_b=\alpha_a=\alpha_c$}
\end{enumerate}
It should be noted that the angle $\theta$ from the above equations is now the angle between the first input wave and the output wave.

\bibliographystyle{jfm}
\bibliography{jfm_paper}

\end{document}